\documentclass[
aip,
 rsi,
 amsmath,amssymb,
preprint,
reprint, 
nofootinbib
]{revtex4-2}

\usepackage{graphicx}
\usepackage{dcolumn}
\usepackage{bm}
\usepackage[mathlines]{lineno}

\usepackage[utf8]{inputenc}
\usepackage[T1]{fontenc}
\usepackage{mathptmx}
\usepackage{etoolbox}
\usepackage{color}
\usepackage{xcolor}
\usepackage{hyperref}
\usepackage{subfigure}

\usepackage{caption}
\usepackage{subcaption}

\makeatletter
\def\@email#1#2{%
 \endgroup
 \patchcmd{\titleblock@produce}
  {\frontmatter@RRAPformat}
  {\frontmatter@RRAPformat{\produce@RRAP{*#1\href{mailto:#2}{#2}}}\frontmatter@RRAPformat}
  {}{}
}%
\makeatother
\begin{document}

 \preprint{LA-UR-26-24570}
\noindent\hfill LA-UR-26-24570

\title{Demonstration of 255-kV high-voltage generation with a Cavallo multiplier system}

\author{S.~M.~Clayton}

\author{T.~M.~Ito}

\author{A.~Jacobs}

\author{A-T.~Le}

\author{M.~F.~Makela}

\author{C.~M.~O'Shaughnessy}

\author{N.~S.~Phan*}
\email[Corresponding author: ]{nphan@lanl.gov}

\author{E.~Renner}

\author{T.~A.~Sandborn}

\author{T.~J.~Schaub}

\author{I.~L.~Smythe}

\author{J.~Surbrook}

\affiliation{Los Alamos National Laboratory, Los Alamos, New Mexico 87545, U.S.A.}

\author{M.~A.~Blatnik}
\affiliation{Kellogg Radiation Laboratory, California Institute of Technology, Pasadena, California 91125, U.S.A.}
\affiliation{Los Alamos National Laboratory, Los Alamos, New Mexico 87545, U.S.A.}

\author{B.~W.~Filippone}
\affiliation{Kellogg Radiation Laboratory, California Institute of Technology, Pasadena, California 91125, U.S.A.}

\date{\today}

\begin{abstract}

Many cryogenic precision measurements require large electric fields in environments where conventional high-voltage feedthroughs are impractical. To address this, we developed a Cavallo electrostatic multiplier designed for \textit{in situ} high-voltage generation under such conditions. Here, we report a room-temperature demonstration of this device. Using a mechanically translated transfer electrode and a custom rotary field mill for noncontact voltage measurement, the system reached output voltages up to approximately $255~\mathrm{kV}$ from a $25~\mathrm{kV}$ DC-biased input voltage in approximately $600~\mathrm{Torr}$ of SF$_6$. The charging curves are quantitatively described by a capacitance-based model once realistic electrode misalignment is included. Voltage-hold measurements show picoampere-scale leakage currents on long time scales, whereas operation near the maximum voltage is limited by transient discharge processes associated with electrode surface condition and local field enhancement, rather than by the intrinsic dielectric strength of the gas. These results demonstrate the Cavallo multiplier as a viable low-current, \textit{in situ} high-voltage source and indicate that electrode surface preparation, alignment tolerances, and insulation performance are the principal requirements for reliable operation in future cryogenic implementations.

\end{abstract}


\maketitle


\section{\label{sec:intro}Introduction}

Many experiments in fundamental physics, including the measurement of electric dipole moments\cite{Baker2010, Ahmed2019} (EDMs) and the search for rare events using large-scale time projection chambers (TPCs), \cite{Rebel2014, nEXO2026} require the application of high voltage to produce intense electric fields in cryogenic environments. However, the use of a high-voltage feedthrough for external power delivery can be impractical in such demanding experimental environments, particularly when multiple operational constraints are present.  In cryogenic apparatuses, this method introduces several challenges, most notably the significant and undesirable heat load generated by leakage currents through the feedthrough\cite{Ito2016}. Additionally, the materials often employed in high-voltage systems can introduce compatibility issues with sensitive experimental instruments, including magnetic contamination in EDM-style measurements and radiopurity constraints in rare-event detectors.\cite{Cianciolo2018, OShaughnessy2013} Moreover, the physical constraints of the experimental apparatus create significant design challenges, which are exacerbated by the magnitude of the high voltage required. A notable example of such a demanding cryogenic high-voltage application that motivated this work is the neutron electric dipole moment (nEDM) search formerly pursued at the Spallation Neutron Source (nEDM@SNS),\cite{Ahmed2019, Ito2026} which required electric fields up to $75~\mathrm{kV/cm}$, corresponding to a potential of $635~\mathrm{kV}$, within a liquid-helium cryostat operating below $0.5~\mathrm{K}$.

\begin{figure*}[htb]
\centering
\includegraphics[width=1.00\linewidth]{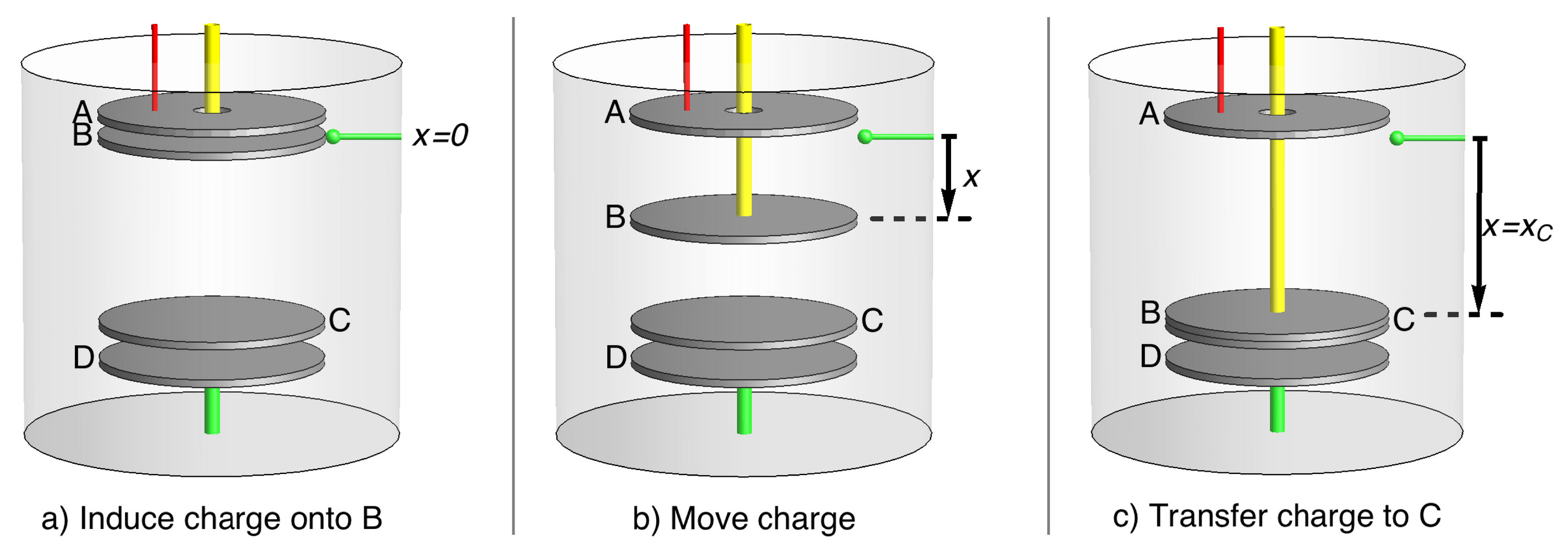}
    \caption{Operating cycle of the Cavallo multiplier. (a) With the transfer electrode $B$ positioned adjacent to the high-voltage bias electrode $A$ and momentarily grounded through the green connection at $x=0$, electrostatic induction drives like charge to ground and leaves $B$ with a net induced charge of opposite polarity. (b) $B$ is disconnected from ground and mechanically translated toward $C$, carrying its isolated charge across the inter-electrode gap. (c) Brought into contact with the isolated collector $C$, supported on an insulating standoff above the grounded electrode ($D$) — $B$ deposits its charge onto $C$; iterating this cycle progressively accumulates charge on $C$, producing a terminal voltage that substantially exceeds the primary voltage applied at $A$. } 
    \label{fig:Cavallo_operation_diagram}
\end{figure*}

To circumvent these limitations, an internal high-voltage generator or multiplier can serve as a viable alternative. In Clayton \textit{et al.},~\cite{Clayton2018} we proposed the use of the Cavallo multiplier,~\cite{Cavallo1795} a classic electrostatic induction machine with mechanical charge transport, for generating high voltage within cryogenic environments. The operating principle of the multiplier is illustrated and described in Fig.~\ref{fig:Cavallo_operation_diagram}. This high-impedance, near-zero-current\footnote{Near-zero-current, or low-current, refers to operation in which the multiplier establishes a large electrostatic potential on a high-impedance load without supplying substantial continuous current.} voltage amplifier is especially well-suited for applications that are highly sensitive to heat loads. We discussed the fundamental operating principle of the Cavallo multiplier, analyzed its maximum voltage gain, and derived the equations for the electrode voltages during the charging phase. To validate our predictions, we presented experimental data from a rudimentary room-temperature demonstrator operated under ambient conditions.

A follow-up study~\cite{Blatnik2026} presented a Cavallo multiplier design specifically customized for the large-scale, cryogenic nEDM experiment~\cite{Ahmed2019, Ito2026} was presented. The electrode geometry is refined using finite-element calculations performed with COMSOL to maximize voltage gain while minimizing the risk of electrical breakdown.~\cite{Phan2021} This design yielded a geometry-defined asymptotic gain of 18, sufficient to step an external input voltage of $50~\mathrm{kV}$ up to the target voltage of $650~\mathrm{kV}$ within a practical number of charging cycles.

In this paper, we report room-temperature operation of an apparatus constructed according to the proposed design. The purpose of this test was to characterize operation and guide iterative refinements before cryogenic operation. However, the dielectric strength of common insulating gases and low-pressure or vacuum insulation in practical systems is substantially lower than that of liquid cryogens, which constrains the maximum achievable voltage to a value below the designed potential of the apparatus.\cite{Mathes1967,Schwenterly1988, Qureshi1995, Christophorou1997} Despite this limitation, the room-temperature measurements provide information on the parameters that govern multiplier performance and therefore inform expectations for operation in other dielectric media. We present data obtained in sulfur hexafluoride (SF$_6$) gas, a dielectric medium commonly employed in high-voltage systems, and discuss key findings essential for ensuring reliable and stable operation of the multiplier.

This paper is organized as follows: Sec.~\ref{sec:setup} provides a description of the experimental apparatus, the methodology for determining the internal system voltage, and the operating procedure. Section~\ref{sec:results} presents the high-voltage multiplication data obtained in SF$_6$. In Sec.~\ref{sec:discussion}, the results are analyzed and compared with theoretical predictions, and implications for future cryogenic operation are discussed. This section also addresses the limitations encountered and discusses challenges related to charge dissipation that require careful consideration. Finally, a summary of the findings is presented in Sec.~\ref{sec:conclusion}.


\section{\label{sec:setup}The apparatus and operating procedure}

\subsection{\label{sec:apparatus}Cavallo multiplier}

\begin{figure*}[htb]
\centering
\includegraphics[width=1.00\linewidth]{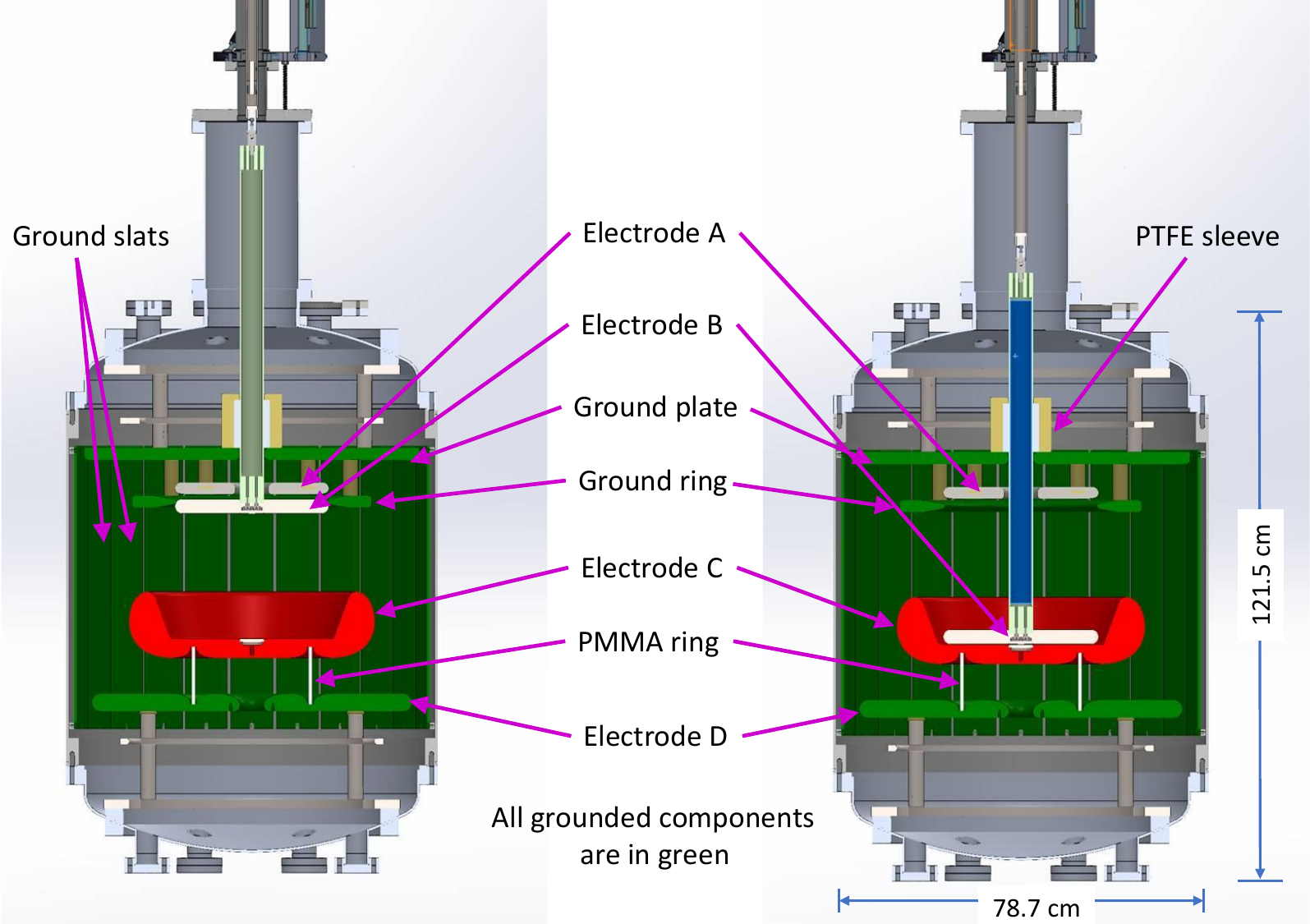}
    \caption{Cross-sectional schematic of the Cavallo multiplier apparatus showing the relative geometry of the primary electrodes and grounded structures (green). For clarity, the external $50$-kV feedthrough, actuator assembly, and other auxiliary hardware are not shown.  The DC-biased electrode $A$, transfer electrode $B$, collector electrode $C$, PMMA support ring, and grounded electrode $D$ are indicated for the two limiting positions of $B$. The figure highlights the bowl-shaped geometry of $C$ and the grounded return boundaries surrounding the multiplier.} \label{fig:CV}
\end{figure*}

\begin{figure*}[htb]
    \centering
	\subfigure[Cavallo bottom]{\includegraphics[width=0.57\linewidth]{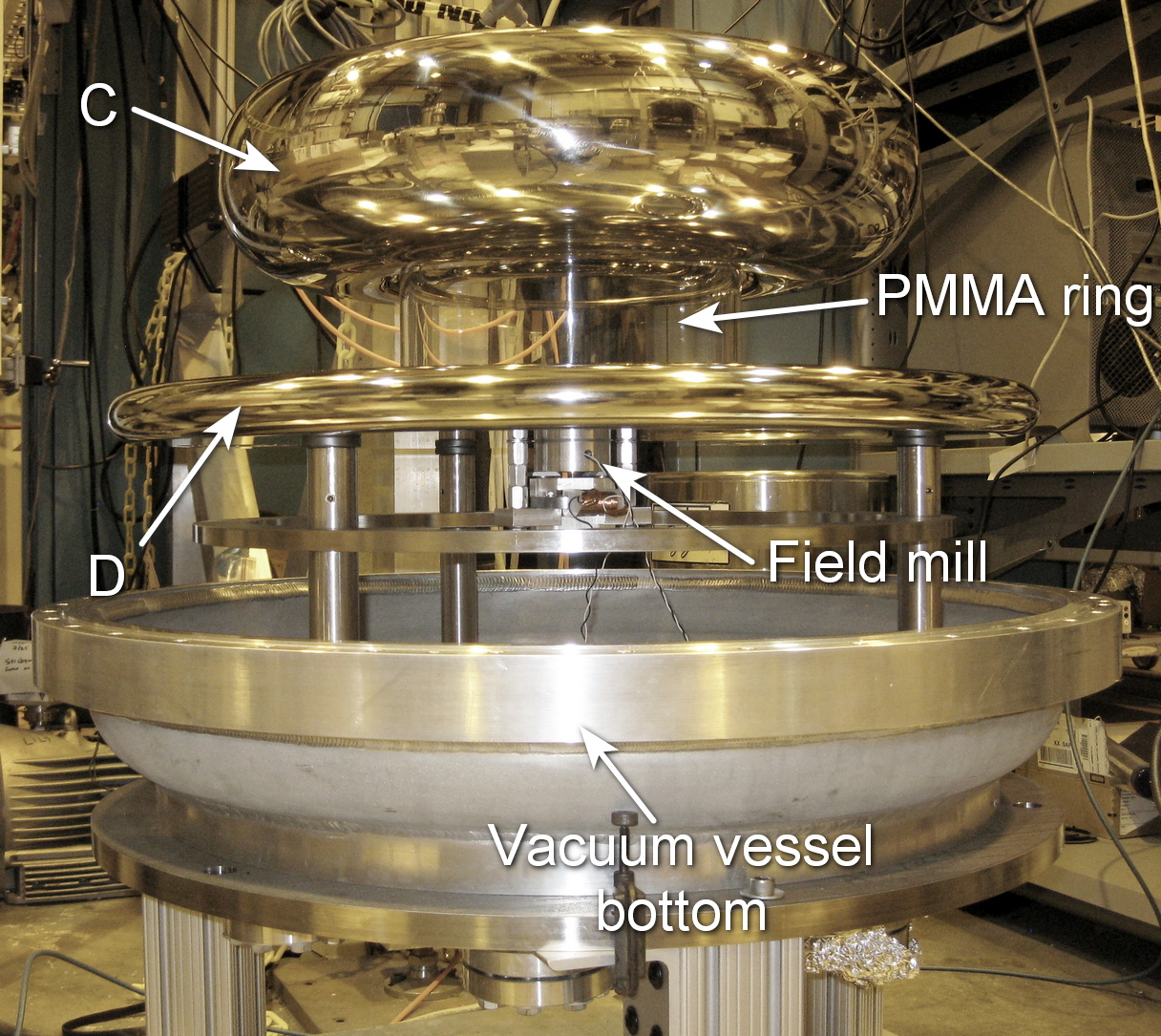}\label{fig:cavallo-bottom}}
    \hspace{0.2cm}
	\subfigure[Ground slats]{\includegraphics[width=0.41\linewidth]{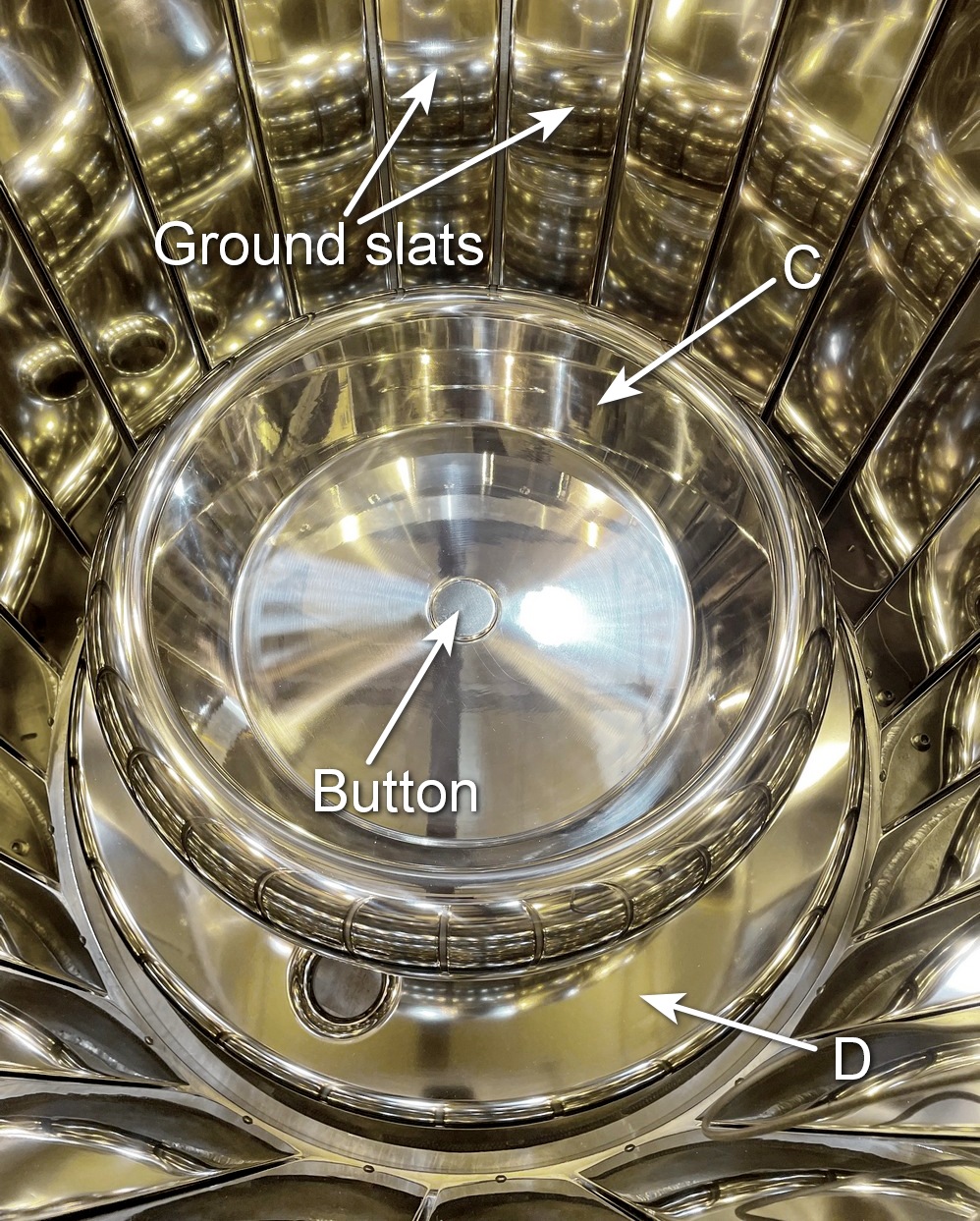}\label{fig:cavallo-slats}}
    \caption{Interior views of the Cavallo multiplier apparatus: (a) the lower portion of the internal assembly with electrode $C$, PMMA ring, electrode $D$, and mounted field mill; (b) the ground slats mounted along the inner wall of the vacuum vessel. Together with the top ground return plate (not shown) and electrode $D$, these slats establish a grounded boundary surface for the system. }
    \label{fig:cavallo-internal}
\end{figure*}

The Cavallo multiplier system, depicted in Fig.~\ref{fig:CV}, consisted of three primary electrodes ($A$, $B$, and $C$) and several grounded electrodes, including a top ground return plate, a ground ring, electrode $D$, and a series of ground return slats (colored green in the figure). Electrode $A$, the uppermost electrode and the `bias electrode', was connected to an external DC high-voltage power supply (Matsusada Precision, Model: AF-100R0.1-LCG) that supplied the input voltage.  The input voltage was delivered through an alumina ceramic feedthrough rated for 50 kV (Ceramtec, Part Number: 21112-01-A), a voltage level compatible with feedthrough technology already demonstrated for cryogenic operation.\cite{Ito2016}  The voltage from the power supply was controlled via a computer, allowing the voltage on $A$ to be held fixed or modulated as necessary. Although the measurements reported here were performed at room temperature, the vessel and its internal hardware were designed from the outset for eventual operation in a cryogenic environment. All materials used in the vessel, electrodes, insulators, support structures, and feedthrough interfaces were selected for cryogenic compatibility in order to preserve mechanical integrity, dimensional stability, and electrical performance during cooldown and thermal cycling. This approach allowed the full apparatus to be tested in a representative configuration before cryogenic operation.

The experimental setup featured a vertically actuated transfer electrode, designated $B$, mounted on a G10 glass-epoxy insulating rod. The rod passed through a bearing assembly mounted on the top flange of the vacuum vessel and through a PTFE guide sleeve attached to the ground plate above electrode~$A$. Its external end was coupled to a linear translator (Thermionics Laboratory, Model: ZC-B450C-7275T-1.71-15). This mechanism enabled the precise translation of electrode $B$ between two specified positions. In the first position, or docked, position electrode $B$ engaged a ground ring around its circumference, with a vertical separation of approximately 5~mm from electrode $A$. A downward extension of the actuator rod brought electrode $B$ to the second position, where it made direct contact with electrode $C$. The electrical connection between electrodes $B$ and $C$ was mediated by a pair of replaceable, 5.08-cm-diameter inserts, or ‘buttons,’ at their centers.  These inserts were designed to be replaceable, which enabled the management of electrode degradation caused by the inherent spark discharge mechanism for charge transfer between electrodes $B$ and $C$ during operation.~\cite{Clayton2018} 

To prevent potential damage from overdriving electrode $B$, a strain gauge (Futek, Model: LCM200) was incorporated into the drive assembly. The output from this gauge served as a limit-sensing mechanism, triggering the actuator to halt when electrode $B$ made physical contact with either the ground ring or electrode $C$. For position monitoring, a laser distance sensor (SICK AG, Model: DT35-B15851) was used to continuously record the position of electrode $B$ during its actuation. The signals from the strain gauge, the optical position sensor, and the preset voltage on electrode $A$ were all routed to a data acquisition system (D-tAcq ACQ2106).

Electrode~$C$ functioned as the high-voltage electrode, or `charge accumulator/collector,' which was raised to the required target voltage of the experiment. In the present apparatus, the effective capacitive load seen by the multiplier, approximately 58~pF, was set primarily by the geometry of electrodes $C$ and $D$. The location of electrode $D$ was chosen so that this load capacitance was approximately representative of that expected from the full high-voltage electrode structure in the intended experiment. Therefore, the present room-temperature configuration provides a realistic test of the charge required to raise the collector to the target high voltage and of the corresponding charging-time behavior, even though it does not reproduce all features of the final cryogenic EDM experiment.

Electrode $C$ featured a bowl-shaped geometry, designed to minimize the parasitic capacitance between electrode~$B$ and ground, $C_{BG}$, thereby maximizing the charge transfer to $C$. It was supported by a 25.4-cm-diameter poly(methyl methacrylate) (PMMA) insulating ring, $12.17~\mathrm{cm}$ high, with a $6.35~\mathrm{mm}$ wall thickness. This ring, in turn, rested upon a grounded electrode, labeled $D$. Both electrodes $C$ and $D$ were precisely machined with smooth grooves to securely accommodate the PMMA ring while simultaneously mitigating the triple-junction effect.\cite{SantAnna2015, Saim2021}  A chamfered hole was centrally located in electrode $D$ for the mounting of a noncontact voltage-monitoring device, referred to as a field mill (shown mounted underneath $D$ in Fig.~\ref{fig:cavallo-bottom}), which is described in the next section. Additionally, three auxiliary holes with similar geometric specifications were machined at equal angular intervals around the perimeter of electrode $D$ to accommodate supplementary instrumentation, such as additional field mills.

The interior wall of the vacuum vessel was lined with 24 ground slats (Fig.~\ref{fig:cavallo-slats}), which together with a top ground return plate, positioned above electrode $A$, and electrode $D$, established a well-defined ground boundary surface for the system.

All electrodes and ground returns within the system were fabricated from stainless steel. Detailed electrode geometries are provided in Blatnik \textit{et al.}.~\cite{Blatnik2026} Following fabrication, the electrodes underwent mechanical polishing. Subsequently, all electrodes and ground returns, with the exception of electrode $C$, were electropolished. The decision not to electropolish electrode $C$ was dictated by practical limitations of the vendor’s equipment, specifically the size of the electrode and the absence of suitable features for current injection required for electropolishing.


\subsection{\label{sec:field-mill}Noncontact voltage measurement: Field mill}

\begin{figure*}
\centering
\includegraphics[width=1.0\linewidth]{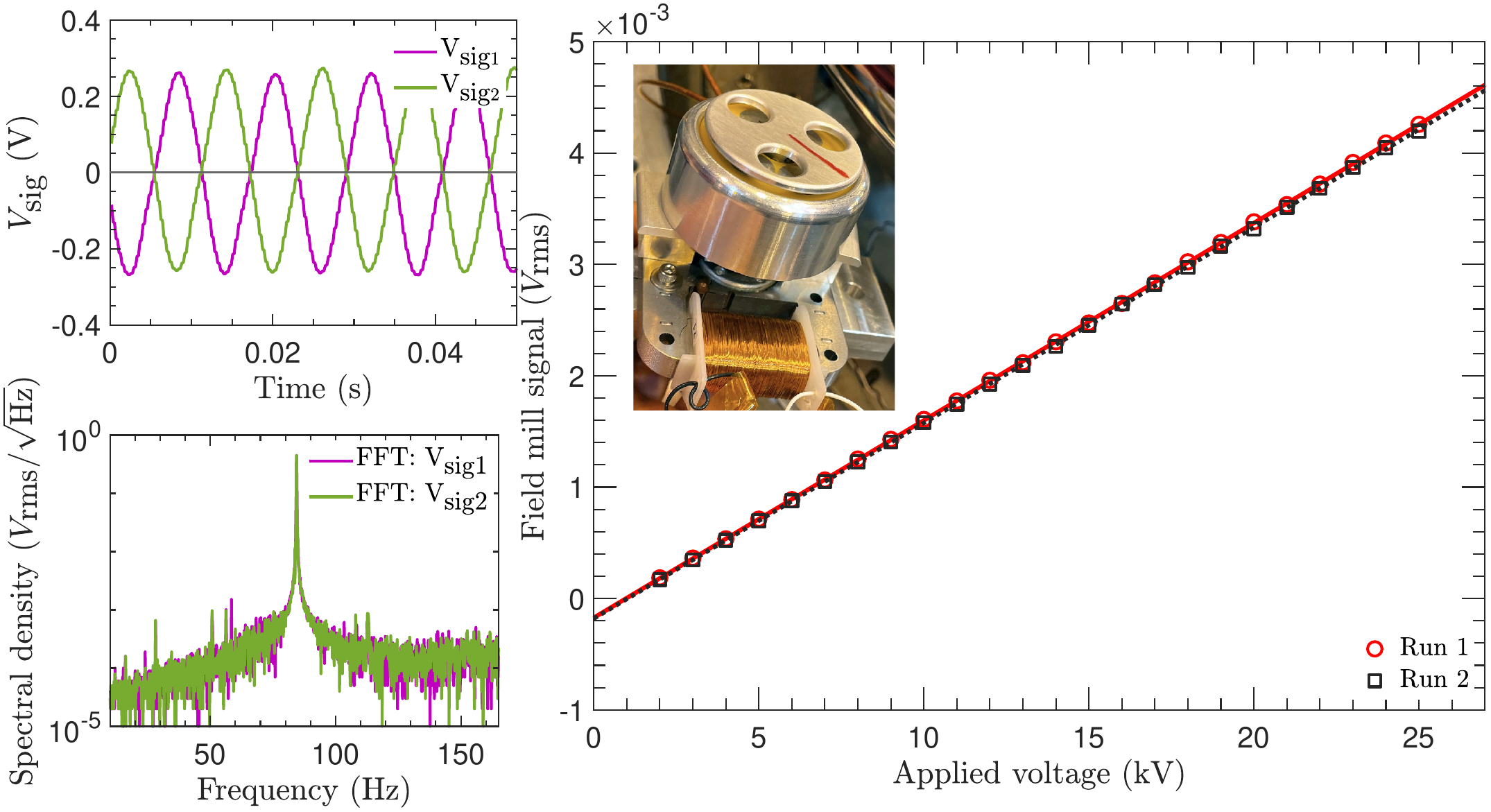}
  \caption{Characterization of the rotary field mill used for noncontact measurement of the voltage on electrode $C$. Top left: representative time-domain signals from the two interleaved sense-pad channels, $V_\mathrm{sig1}$ and $V_\mathrm{sig2}$. Bottom left: corresponding Fourier spectra, with a dominant peak at $84.4~\mathrm{Hz}$, the field-mill rotation frequency. Right: photograph of the field mill and calibration curve of its output, expressed as RMS signal amplitude, versus the applied voltage on electrode $C$ for two independent runs; the linear response demonstrates the suitability of the device for noncontact determination of the $C$ electrode voltage.}
    \label{fig:field-mill-tile-plot}
\end{figure*}

Because electrode $C$ is electrically isolated and serves as a high-impedance charge collector, its potential must be determined without introducing a direct conductive measurement path that would perturb the accumulated charge. Conventional contact-based diagnostics, such as resistive dividers or probe connections, would introduce leakage current, parasitic loading, and additional hardware constraints that could alter the true operating conditions of the multiplier. Therefore, a noncontact technique is required to monitor the collector voltage while preserving galvanic isolation and the intrinsic high-impedance character of the apparatus. This approach is especially important for enclosed or cryogenic high-voltage systems, in which direct electrical access to the energized electrode is often impractical or undesirable.

The voltage on the $C$ electrode was measured using a rotary field mill,\cite{Hollandsworth1963,Li2024} a device commonly employed for noncontact high-voltage measurement on energized sources.\cite{Nicoll2019, Handel2022} This technique operates by converting a static electric field into a quantifiable alternating current (AC) signal. The core mechanism involves a grounded, rotating shutter that periodically modulates the exposure of a sensor plate to the external electric field. When an aperture in the rotor aligns with the sensor plate, the external electric field induces charge on the plate. Conversely, when a shield on the rotor aligns with the sensor plate, the plate is screened from the external electric field.  This cyclical exposure and shielding of the sensor plate generates a small, oscillating AC current (shown in the upper-left panel of Fig.~\ref{fig:field-mill-tile-plot}). The amplitude of the resulting AC signal is directly proportional to the intensity of the static electric field being measured. Following calibration, this output can be used to determine the potential of the high-voltage electrode.

The position of the custom-built field mill used to measure the high voltage generated by the Cavallo apparatus is shown in Fig.~\ref{fig:cavallo-bottom}. The field mill was driven by a small C-frame shaded-pole induction motor, a single-phase AC motor well-suited for light-load continuous operation (inset of the right panel of Fig.~\ref{fig:field-mill-tile-plot}). This motor type was selected after preliminary testing that confirmed its compatibility with both vacuum and cryogenic environments. The design and characterization of the field mill will be reported separately.

The field mill housing incorporated a precision-machined lip to accommodate the printed circuit board (PCB) sense pads. This design feature ensured that the six radially symmetric pads were positioned perpendicular to the shaft of the rotating vane and set a fixed gap of approximately 6 mm between the vane and the sense pads. For compatibility with cryogenic and vacuum operation, two full-ceramic R144 bearings were selected. A spring-and-plug assembly was integrated to ground the shaft while minimizing rotational friction.

To mitigate corona discharge, the vane was designed with internal holes to minimize sharp edges. The risk of corona discharge was further reduced by recessing the entire field mill assembly into the central aperture of the $D$ electrode, as illustrated in Figs.~\ref{fig:CV} and \ref{fig:cavallo-bottom}.

The output from the field-mill sense electrodes was routed out of the apparatus and connected to a low-noise current preamplifier (Stanford Research Systems 570). The amplified output signal was subsequently digitized by a D-tAcq data acquisition system (ACQ2106) for processing and analysis.


\subsection{\label{sec:procedure}Procedure}

\subsubsection{\label{sec:calibration}Voltage calibration}

To calibrate the field mill signal relative to the voltage on the $C$ electrode, known voltages were applied to the $C$ electrode with the system immersed in the target gas. This procedure was facilitated by a magnetically actuated grounding rod, which was installed on one of the top ports of the vacuum chamber. This rod served a dual purpose: it could be lowered to make contact with the lobe of the $C$ electrode (i.e. the top rim of the bowl) to ground and de-energize the system, or it could be electrically isolated from ground and connected to an external high-voltage power supply. The right panel in Fig.~\ref{fig:field-mill-tile-plot} shows the resulting linear relationship between the applied voltages on the $C$ electrode and the corresponding output measured by the field mill.

The calibration curves were cross-checked by integrating the charge drained from electrode~$C$ with an electrometer during the calibration procedure. This provided an independent consistency check of the voltage inferred from the field-mill response over the directly calibrated range shown in Fig.~\ref{fig:field-mill-tile-plot}. Within the precision of the measurement, the electrometer-based estimate was consistent with the linear field-mill calibration, although a systematic offset remained evident as a nonzero intercept that varied between runs.

It is important to note that the electrometer measurement was sensitive only to the magnitude of the drained charge and did not determine its sign.  The offset variation was believed to result from charge accumulation on the PMMA ring separating electrodes $C$ and $D$. To mitigate this effect, an ionizer was introduced to neutralize stray charges within the experimental volume. This procedure reduced the calibration intercept by approximately 50\%, but did not eliminate the offset. The $y$-intercept subsequently increased again following the deactivation of the ionizer, further supporting the hypothesis of stray charge accumulation.\cite{Fakhfakh2012}

The field-mill response was found to be linear and reproducible over the directly calibrated range shown in Fig.~\ref{fig:field-mill-tile-plot}. The residuals about the best-fit linear calibration for the two runs show no systematic curvature. Over the directly calibrated range of $2$--$25~\mathrm{kV}$, the maximum deviation from proportionality is less than $0.1~\mathrm{kV}$ (less than $0.4\%$ of full scale) in either run.  For the directly calibrated polarity, that is, positive $C$ electrode voltage corresponding to negative bias voltage on electrode~$A$, the statistical uncertainty in the inferred $C$-electrode voltage was obtained by propagating the fit uncertainties in the calibration slope and intercept. Evaluated at a reference voltage of $V_C=255~\mathrm{kV}$, chosen for comparison with the maximum inferred collector voltage reported below, this gives $\pm0.7~\mathrm{kV}$, assuming the linear calibration remains valid up to the inferred voltage range. A separate systematic uncertainty was estimated from the run-to-run variation of the two independent calibration curves shown in Fig.~\ref{fig:field-mill-tile-plot}. At the same reference voltage, this calibration-to-calibration variation corresponds to $\pm1.6~\mathrm{kV}$ on a $1\sigma$ basis, while the full spread between the two calibration curves corresponds to $\pm2.3~\mathrm{kV}$ and may be taken as a conservative bound on the absolute voltage scale. Unless otherwise noted, voltages quoted for the directly calibrated polarity use the combined $1\sigma$ calibration uncertainty, obtained by adding the statistical and calibration-to-calibration terms in quadrature. At $V_C=255~\mathrm{kV}$, this gives $\pm1.7~\mathrm{kV}$. Because the calibration is linear, these uncertainty estimates scale approximately in proportion to $V_C$. All direct calibration curves shown in Fig.~\ref{fig:field-mill-tile-plot} were acquired for positive $C$-electrode voltage. The opposite $C$-electrode-voltage polarity was inferred using the same linear calibration, but was not independently established; it is examined in more detail in Sec.~\ref{sec:discuss-comparision}.


\subsubsection{Charging process}\label{sec:charging}

Electrode~$C$ was charged by holding electrode~$A$ at a fixed potential (e.g., $V_A = 10~\mathrm{kV}$) while electrode~$B$ was cycled between its grounded, docked position near $A$ and its contact position at $C$. For concreteness, the description below assumes $V_A>0$; the same sequence applies for opposite polarity with the charge signs reversed. In each cycle, charge is induced on $B$ while it is grounded near $A$, mechanically transported after isolation from ground, and then deposited onto the floating collector electrode~$C$. A diagram illustrating this process is shown in Fig.~\ref{fig:Cavallo_operation_diagram}.

\begin{enumerate}
    \item \textbf{Induction and grounding:} With $B$ docked near the positively biased electrode~$A$ and connected to the ground ring, electrons are drawn onto $B$ while positive charge flows to ground, leaving $B$ with a net negative charge.

    \item \textbf{Isolation and transport:} Electrode~$B$ is moved away from $A$, breaking the ground connection, and carries the induced charge $Q_B$ toward electrode~$C$.

    \item \textbf{Charge transfer:} When $B$ contacts the floating electrode~$C$, charge redistributes over the combined $B$--$C$ conductor and a portion is transferred to $C$. The nearby grounded electrode~$D$ enhances this transfer through capacitive coupling, lowering the potential of $C$ and inducing further charge deposition.

    \item \textbf{Return and repetition:} Electrode~$B$ is withdrawn from $C$ and returned to its docked position near $A$, where the cycle is repeated until the desired voltage is reached or the system approaches saturation.
\end{enumerate}

With each cycle, additional charge is drawn from ground via induction and added to electrode $C$.  After a sufficient number of cycles, $N$, electrode~$C$ develops a very high electrostatic potential relative to ground. The achievable final voltage either approaches the gain-limited saturation value of the system or is limited earlier by leakage through the surrounding medium or by spark breakdown.

If saturation is reached, its magnitude is governed by the voltage on electrode~$A$, $V_A$, and the mutual capacitances $C_{ij}^{z}$ of the electrodes in the system. Following Clayton \textit{et al.},~\cite{Clayton2018} the saturation voltage is
\begin{equation}
\label{eq:Vcmax}
\begin{aligned}
V_C^{\max}=-V_A\frac{C_{AB}^{a}-C_{AB}^{c}}{C_{BG}^{c}+C_{AB}^{c}+\kappa C_{BC}^{a}},
\end{aligned}
\end{equation}
where $\kappa  = C_{\mathrm{CG}}^{c}/ (C_{\mathrm{CG}}^{a} + C_{BC}^{a}) \approx 1$.  The subscripts $ij$ denote the electrode pair while the superscript $z$ specifies the position of electrode $B$ ($z=a$: electrode $B$ next to electrode $A$ and docked at the ground ring; $z=c$: electrode $B$ contacting electrode $C$). Together, these quantities determine the voltage gain factor of the multiplier.  A higher saturation potential is achieved by performing the charging process with a higher voltage applied to electrode $A$. 

To mitigate sparking and suppress premature discharge during re-docking of electrode $B$ at the grounded ring, the charging procedure was modified so that the voltage on electrode $A$ was temporarily reduced during the return stroke of $B$. This empirical voltage-adjustment step lowered the electric field between $A$ and $B$ while $B$ still carried residual charge, thereby helping to equilibrate the electrode potentials and minimize unwanted charge transfer or discharge when $B$ contacted the ground ring. After re-docking, the voltage on $A$ was restored to its nominal value and the next charging cycle was initiated. This modified operating procedure was found to produce smoother, more reproducible charging curves.

In the present measurements, one complete actuation cycle of electrode~$B$---from the grounded docked position near electrode~$A$, to contact with electrode~$C$, and back---required approximately $39~\mathrm{s}$. This cycle period defines the characteristic step spacing in the charging curves shown below. More generally, the cycle rate cannot be increased arbitrarily in a cryogenic implementation, because the allowable speed of electrode~$B$ is constrained not only by mechanical considerations but also by the heat load associated with electrode motion and charge transfer, as discussed by Clayton~\textit{et al}.\cite{Clayton2018}


\section{Experimental results}\label{sec:results}

\begin{figure*}
    \centering
	{\includegraphics[width=\linewidth]{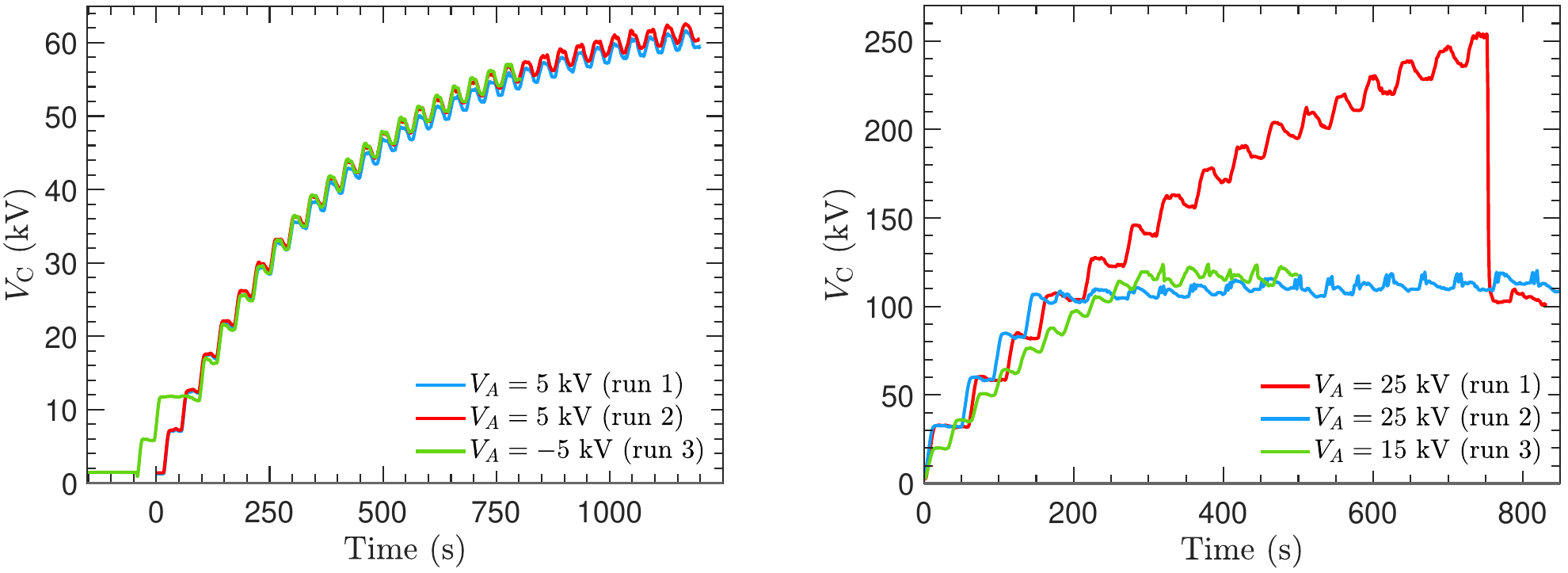}}
    \caption{Charging curves of the Cavallo multiplier measured in SF$_6$. Left: electrode $C$ voltage $V_C$ as a function of time for both positive and negative polarities applied to electrode $A$ at $V_A=\pm5$~kV, showing similar stepwise charging behavior for both bias polarities. Right: charging behavior at higher bias voltage, illustrating the effect of discharge-induced surface degradation. An initial run with $V_A=25$~kV reaches $V_C\approx254.6$~kV, whereas subsequent runs are limited to a lower voltage ceiling of approximately 110–120~kV.}
    \label{fig:charging-curves-1}
\end{figure*}

\begin{figure*}[htb]
    \centering
	{\includegraphics[width=\linewidth]{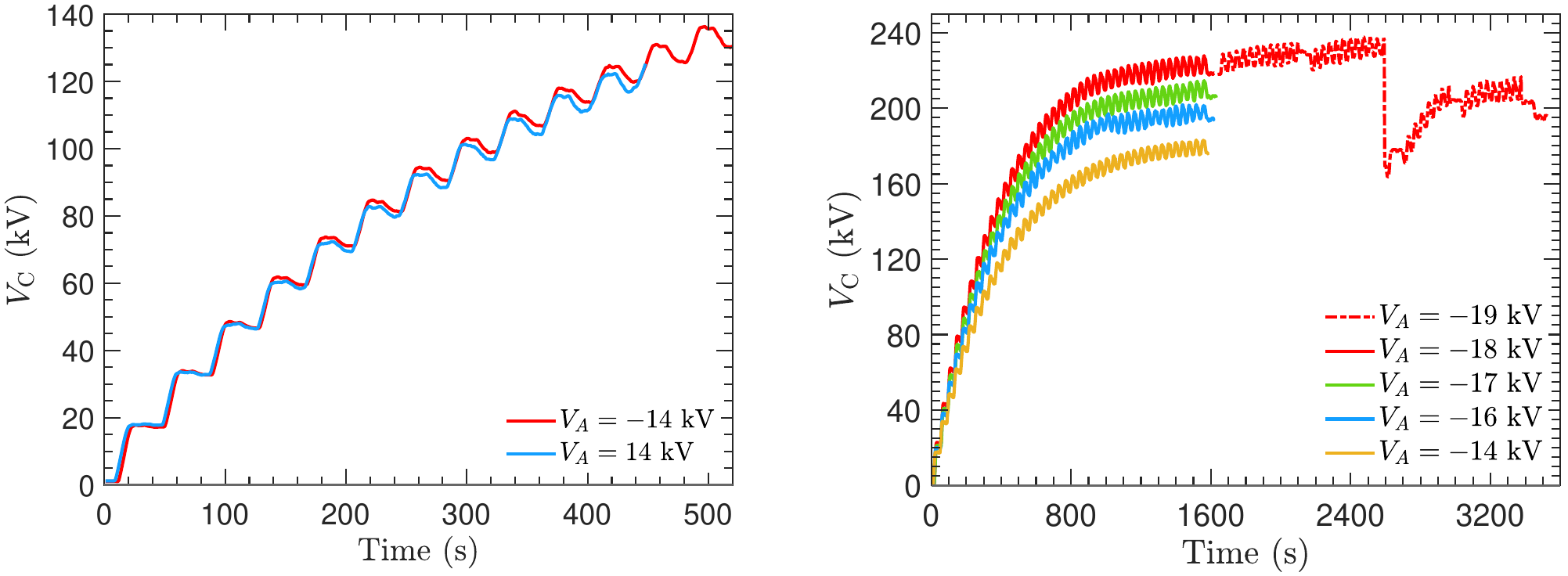}}
    \caption{Charging curves of the Cavallo multiplier in SF$_6$. Left: electrode $C$ voltage as a function of time for both positive and negative polarities applied to electrode $A$ at $V_A=\pm14$ kV, demonstrating similar stepwise charging behavior for the two polarities for larger bias voltages. Right: $V_C(t)$ for several negative values of $V_A$, acquired after repolishing the damaged regions of the ground ring and electrode~C following the discharge event shown in Fig.~\ref{fig:charging-curves-1}. The charging rate and the maximum voltage both increase with increasing $|V_A|$, with the highest-voltage runs again exhibiting the onset of greater instability near saturation.}
    \label{fig:charging-curves-2}
\end{figure*}

Initial system commissioning was conducted under vacuum to verify mechanical and electrical reliability and to confirm that the system functioned according to design specifications. Because the system was designed primarily for operation in cryogenic fluids, the achievable vacuum was limited to a base pressure of approximately $4\times10^{-5}$~Torr due to a combination of outgassing, especially from plastic and fiberglass components, and likely virtual leaks. Consequently, performance testing was conducted within a controlled atmosphere of sulfur hexafluoride (SF$_6$) at approximately 600 Torr. This medium was selected to enable higher operating voltages than those attainable under the suboptimal vacuum conditions.\cite{Malik1979} Although SF$_6$ was subsequently introduced for performance characterization, the underlying feasibility of operating the system under vacuum was confirmed during the initial verification stage.

Figures~\ref{fig:charging-curves-1} and \ref{fig:charging-curves-2} summarize the charging behavior measured in SF$_6$. The voltage on electrode $C$ exhibits the expected stepwise rise associated with successive transfer cycles of electrode $B$, superimposed on an overall saturating envelope. For moderate values of the bias voltage, similar charging behavior is observed for both polarities applied to electrode $A$, indicating approximate polarity symmetry of the charging dynamics. As $|V_A|$ is increased, both the charging rate and the maximum value of $V_C$ increase, while the highest-voltage runs show the onset of greater instability near saturation.  As discussed in Sec.~\ref{sec:discuss-comparision}, however, quantitative comparison of the absolute $C$ electrode voltage between the two polarities is limited by a sign-dependent calibration effect in the field mill.

In SF$_6$, the maximum field-mill-inferred $C$-electrode voltage observed in the present study was approximately $255~\mathrm{kV}$ for $V_A=25~\mathrm{kV}$, after which a major discharge occurred at approximately $750~\mathrm{s}$ (right panel of Fig.~\ref{fig:charging-curves-1}). Because the direct field-mill calibration was not established for this $C$-electrode-voltage polarity, this value carries an additional sign-dependent systematic uncertainty.  As discussed in Sec.~\ref{sec:discuss-comparision}, the systematic calibration uncertainty is estimated to be about $5.8~\mathrm{kV}$ at the maximum inferred voltage of 255~kV, beyond the combined $1\sigma$ calibration uncertainty defined in Sec.~\ref{sec:calibration} for the directly calibrated polarity. Following this event, subsequent positive-polarity charging cycles, using the initial $V_A=25~\mathrm{kV}$ and a reduced $V_A=15~\mathrm{kV}$, were limited to approximately $110$--$120~\mathrm{kV}$ and exhibited markedly reduced stability until the affected regions of the ground ring and electrode~$C$ were mechanically repolished (1000- and 2000-grit sandpaper). A post-test examination of the electrodes revealed burn marks on both the ground ring and electrode~$C$. After repolishing, stable high-voltage charging behavior was recovered, as shown in the right panel of Fig.~\ref{fig:charging-curves-2}. The recovery of high-voltage performance after repolishing is therefore consistent with the conclusion that the attainable voltage in SF$_6$ is strongly dependent on the post-discharge surface condition of the electrodes and on the resulting local field enhancement.

For the directly calibrated polarity, corresponding to negative bias voltage on electrode~$A$, the highest measured collector voltage in the present study was $V_C=(237.7\pm1.7)~\mathrm{kV}$ at $V_A=-19~\mathrm{kV}$. This represents the highest voltage reported here whose absolute scale is established without the additional polarity-dependent calibration systematic discussed in Sec.~\ref{sec:discuss-comparision}.

Regarding post-discharge behavior, one possible mechanism is discharge-induced chemical modification of the electrode surfaces in SF$_6$. Although pure SF$_6$ is highly inert, an electrical discharge can cause its decomposition into reactive and corrosive byproducts.\cite{Schumb1949} The high temperatures within a discharge break the S--F bonds, producing lower-order sulfur fluorides (e.g., SF$_4$, S$_2$F$_{10}$) and atomic fluorine.\cite{Tsai2007} These decomposition products can react with the stainless-steel electrodes to form solid metal fluorides, and in the presence of trace moisture (H$_2$O) can also lead to the formation of hydrofluoric acid (HF).\cite{Yang2021, Tang2017} Such processes could alter the surface topography of the electrodes, creating new microscopic irregularities and thereby lowering the breakdown voltage of the system. However, no post-run chemical or surface-composition analysis was performed in the present study, so this mechanism should be regarded as an interpretation consistent with the observed damage and performance recovery, rather than a verified hypothesis.


\section{Discussion}\label{sec:discussion}

\subsection{Comparison with the capacitance model}\label{sec:discuss-comparision}

\begin{figure*}[htb]
\centering
\includegraphics[width=1.00\linewidth]{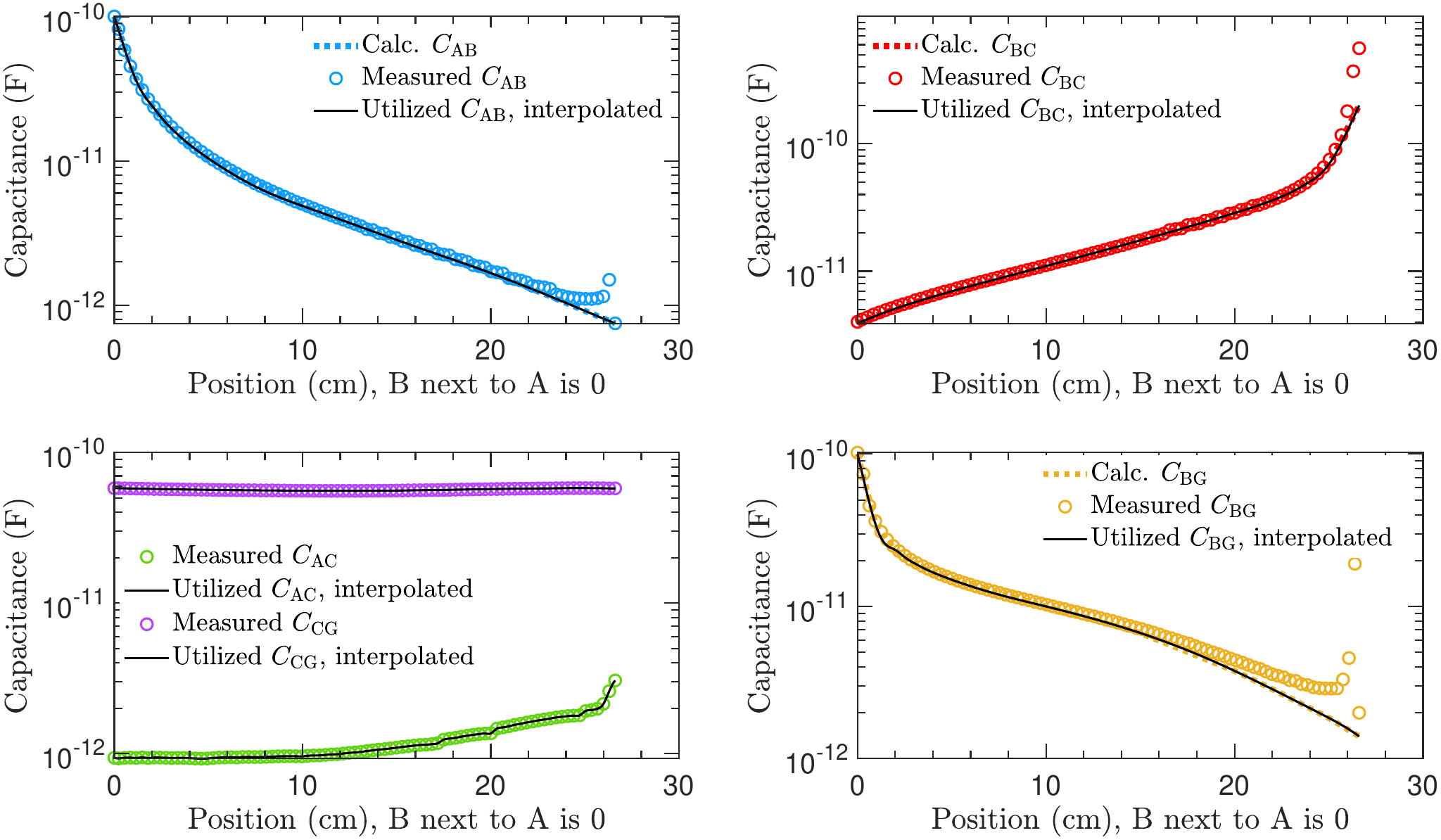}
    \caption{Measured and calculated capacitances between relevant electrode pairs as functions of the vertical position of the transfer electrode~$B$. The panels show $C_{\mathrm{AB}}$ (top-left), $C_{\mathrm{BC}}$ (top-right), $C_{\mathrm{AC}}$ and $C_{\mathrm{CG}}$ (bottom-left), and $C_{\mathrm{BG}}$ (bottom-right). Symbols denote measured values, dashed curves denote the electrostatic calculations, and solid curves denote the interpolated capacitances used in the charging model. The horizontal axis is defined such that $z_B = 0$ corresponds to electrode~$B$ adjacent to electrode~$A$, while $z_B = 26.6~\mathrm{cm}$ corresponds to the position of electrode~$B$ at electrode~$C$; thus quantities such as $C_{\mathrm{AB}}^{a}$ and $C_{\mathrm{AB}}^{c}$ denote the endpoint values at these two positions.} \label{fig:capacitances}
\end{figure*}

\begin{figure*}[htb]
    \centering
    \subfigure[]{\includegraphics[width=0.47\textwidth]{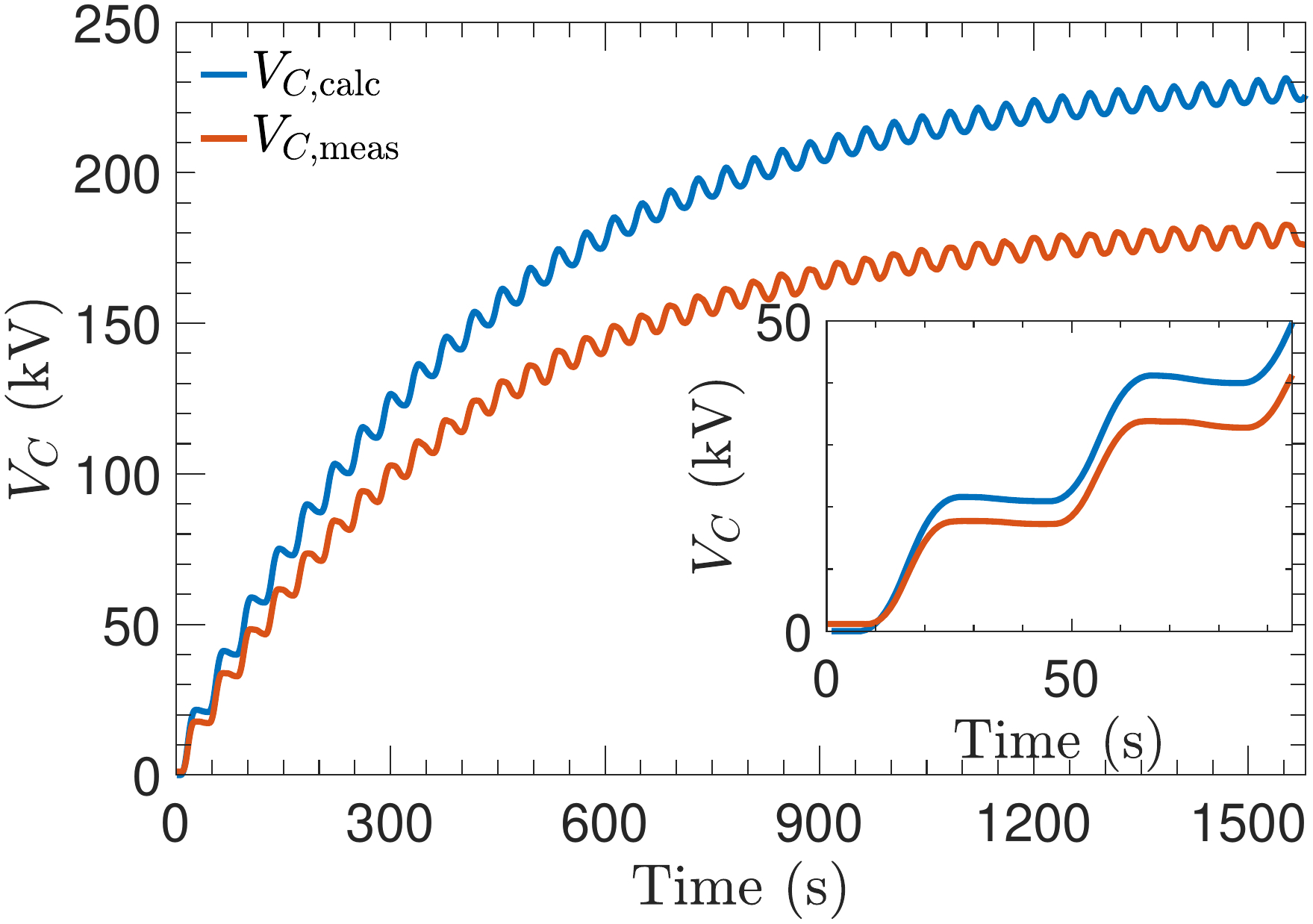}\label{fig:charging_curve_neg14kV_capacitance_model_a}} 
    \hspace{0.25cm}
    \subfigure[]{\includegraphics[width=0.47\textwidth]{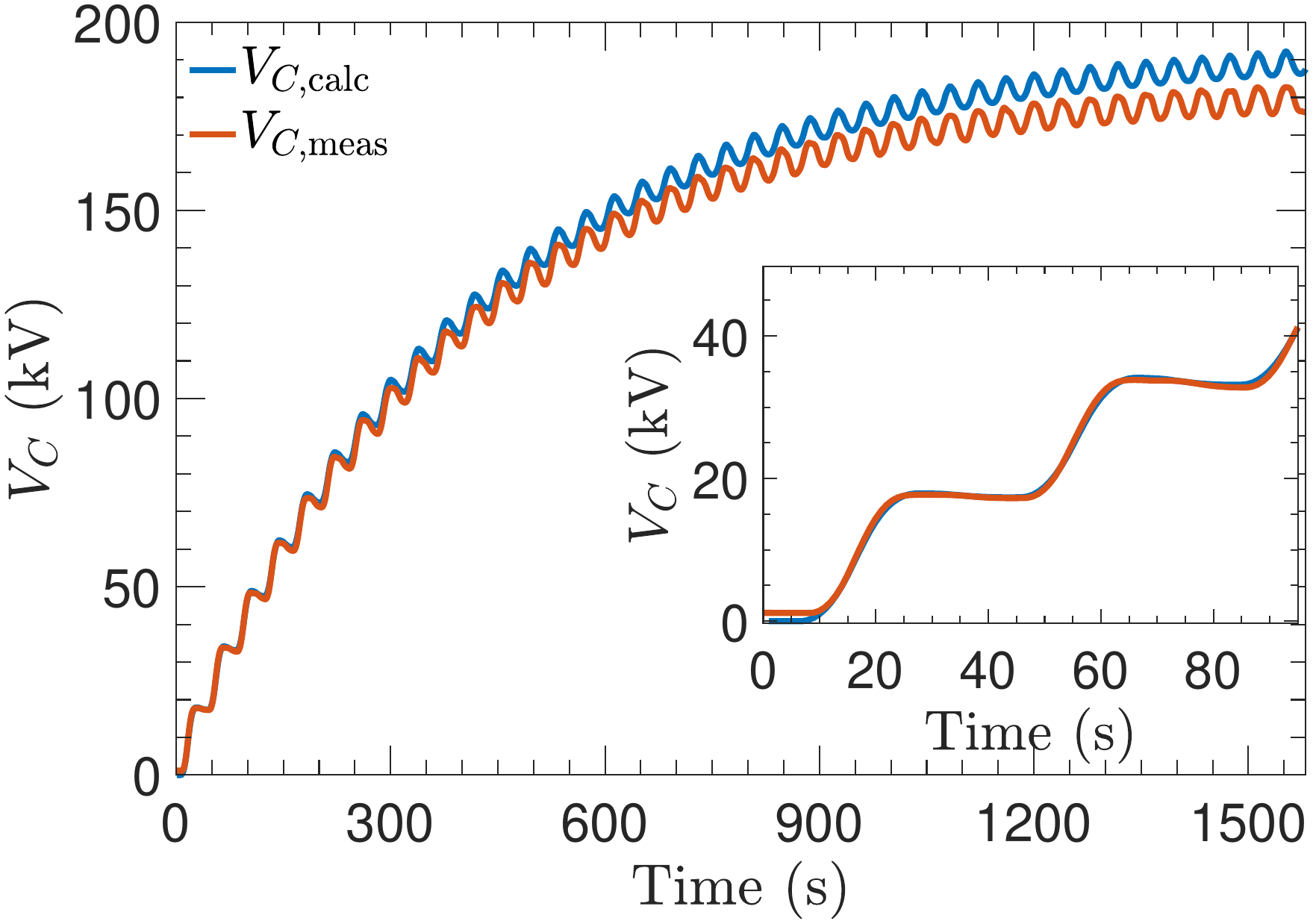}\label{fig:charging_curve_neg14kV_capacitance_model_b}} 
    \subfigure[]{\includegraphics[width=0.47\textwidth]{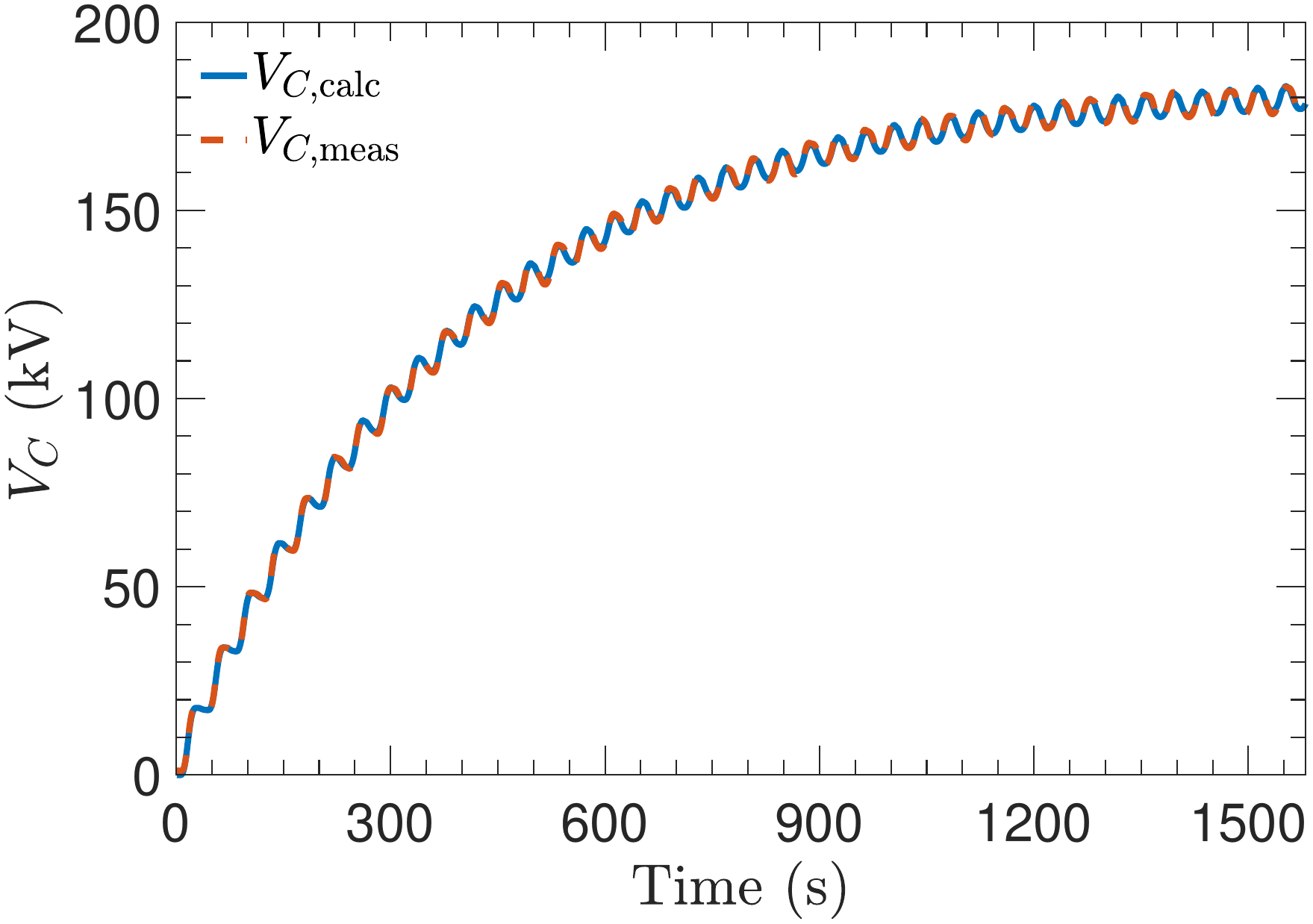}\label{fig:charging_curve_neg14kV_capacitance_model_c}}
    \hspace{0.25cm}
    \subfigure[]{\includegraphics[width=0.47\textwidth]{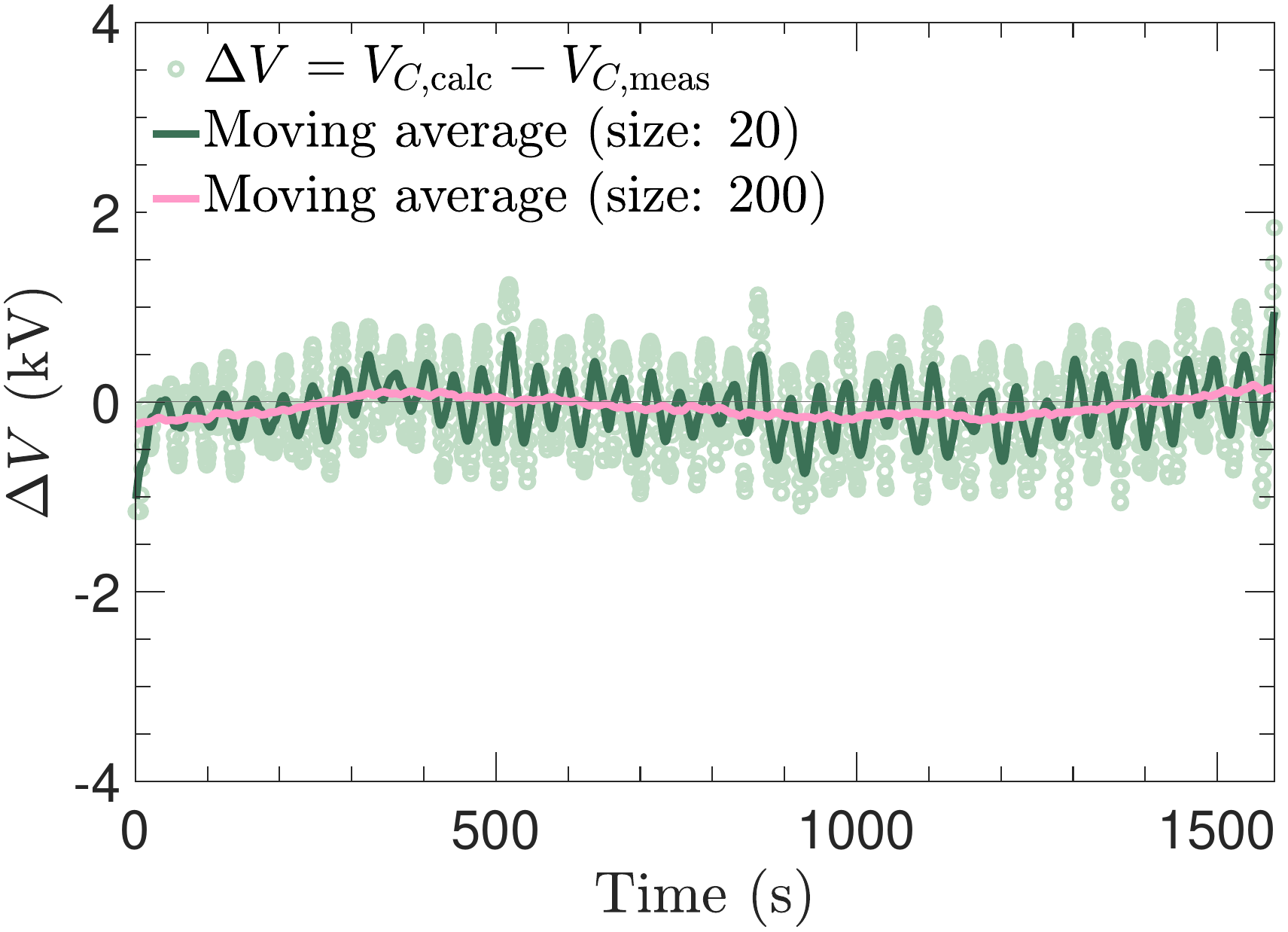}\label{fig:charging_curve_neg14kV_capacitance_model_d}}
    \caption{Comparison of measured and calculated charging curves for $V_A=-14~\mathrm{kV}$. (a) $C$ electrode voltage $V_C(t)$ obtained from the nominal capacitance model and from the measurement. (b) Model with $C_{AB}^{a}$ reduced by $17\%$. (c) Model with $C_{AB}^{a}$ reduced by $17\%$ and $C_{BG}^{c}$ increased by $24\%$. (d) Voltage difference $\Delta V = V_{C,\mathrm{calc}} - V_{C,\mathrm{meas}}$ for the final adjusted model, with moving averages indicating good overall agreement.}
\label{fig:charging_curve_neg14kV_capacitance_model}
\end{figure*}

The comparison with the capacitance model is organized to separate three effects that can produce differences between the measured and calculated charging curves. The first is a voltage-independent geometric effect, in which small offsets or tilts of the electrodes modify the endpoint capacitances that determine charge induction and charge transfer. The second is a voltage-dependent loss mechanism that becomes important only after the $C$ electrode reaches a sufficiently high field. The third is a polarity-dependent field-mill calibration effect, which affects the inferred absolute voltage scale for the polarity not directly established by the calibration. Therefore, we begin with the moderate-voltage ($V_A=-14~\mathrm{kV}$) data to assess whether realistic geometric corrections are sufficient to reproduce the observed charging curve.  We then apply the same capacitance corrections to the higher-voltage data to identify the onset of additional field-dependent losses.  Finally, we compare the two polarities to quantify the remaining calibration asymmetry.

To carry out this comparison, we construct theoretical charging curves from the position-dependent inter-electrode capacitances and the measured actuation sequence.  The relevant capacitances, obtained from both finite-element calculations performed with COMSOL and from direct measurement, are shown in Fig.~\ref{fig:capacitances}. Overall, the agreement between the calculated and measured capacitances is good over most of the electrode travel, which supports the use of the capacitance model as a quantitative description of the apparatus. The endpoint positions exhibit the largest discrepancies because the geometry is highly sensitive to alignment at these locations and parasitic effects affect measurement accuracy. Therefore, in the analysis below, we take the calculated capacitances as the baseline inputs for the model charging curves.

The comparison is especially sensitive to the capacitances that govern charge induction on electrode~$B$ near electrode~$A$ and charge transfer from $B$ to $C$. According to Eq.~\ref{eq:Vcmax}, the saturation voltage depends strongly on $C_{AB}^{a}$, since $C_{AB}^{c} \ll C_{AB}^{a}$ and the induced charge on the transfer electrode is therefore largely set by the former quantity. The denominator of Eq.~\ref{eq:Vcmax} contains the terms $C_{BG}^{c}$, $C_{AB}^{c}$, and $\kappa C_{BC}^{a}$, which are of comparable order, but $C_{BG}^{c}$ is expected to be the most sensitive to mechanical misalignment because it depends on the relative position of $B$ within the bowl-shaped geometry of electrode~$C$. Because the nominal gap between electrodes $A$ and $B$ is only $5~\mathrm{mm}$, positional offsets at the level of a few hundred micrometers can measurably affect the charge induced on $B$ and, consequently, the overall charging efficiency of the multiplier.

This sensitivity is evident in the comparison for $V_A=-14~\mathrm{kV}$ shown in Fig.~\ref{fig:charging_curve_neg14kV_capacitance_model}. Using the nominal capacitances, the calculated charging curve overestimates the measured $C$ electrode voltage beginning with the first charging cycle, and the discrepancy grows with successive cycles. The immediate disagreement in the first few charging steps indicates that the amount of charge induced on electrode~$B$ is smaller than predicted by the ideal geometry. A substantially improved description of the early-time behavior is obtained by reducing $C_{AB}^{a}$ by $17\%$, which is consistent with a small misalignment between electrodes~$A$ and~$B$. Although the present capacitance analysis does not determine the corresponding mechanical displacement, a correction of this magnitude is compatible with a modest offset or tilt of electrode~$B$ at a level that is reasonable for practical assembly tolerances of the apparatus. However, although this adjustment reproduces the first charging cycles, a systematic divergence remains at later times, indicating that reduced induction alone is not sufficient to explain the full difference between the measured and calculated curves.

The remaining discrepancy can naturally be explained by an effective lower-position coupling of electrode $B$ to ground that is larger than predicted by the nominal geometry. The measured and calculated $C_{BG}(z_B)$ curves begin to diverge before electrode~$B$ reaches the final $C$-contact position (bottom-right plot of Fig.~\ref{fig:capacitances}), indicating that the effect is not confined to the instant of $B$--$C$ contact but develops as $B$ enters the lower $C$/$D$ region. Because $C_{BG}$ is itself only of order a few picofarads, even modest stray capacitances or small changes in the local electrostatic boundary conditions can produce a substantial fractional shift. One possible contributor is triboelectric charge accumulation on the G10 actuator rod as it slides through the PTFE guide sleeve, which may alter the local electrostatic boundary conditions in a motion-dependent manner and thus contribute to an apparent increase in the effective lower-position coupling of $B$ to ground.  This behavior is also consistent with a modest off-axis displacement, tilt, or other geometric deviation of $B$ relative to the bowl-shaped $C$ electrode, or with additional parasitic coupling to nearby grounded structures not fully captured by the nominal electrostatic model. In this regard, the fitted increase in $C_{BG}^{c}$ by $24\%$, which yields the agreement shown in Fig.~\ref{fig:charging_curve_neg14kV_capacitance_model_c}, represents an \textit{effective} lower-position correction due to the combined contributions of those effects rather than a single mechanical offset.  COMSOL calculations of misaligned configurations of electrode~$B$ show that small angular tilts of $B$ relative to $A$ of only 0.5$^{\circ}$ can lead to a reduction of 17\% in $C_{AB}^{a}$ and an $8\%$ increase in $C_{BG}^{c}$. Although the present analysis does not attempt to reconstruct the underlying mechanical misalignment, these calculations show that the capacitance corrections used in the model are consistent with practical assembly tolerances.

\begin{figure*}[htb]
\centering
\includegraphics[width=1.00\linewidth]{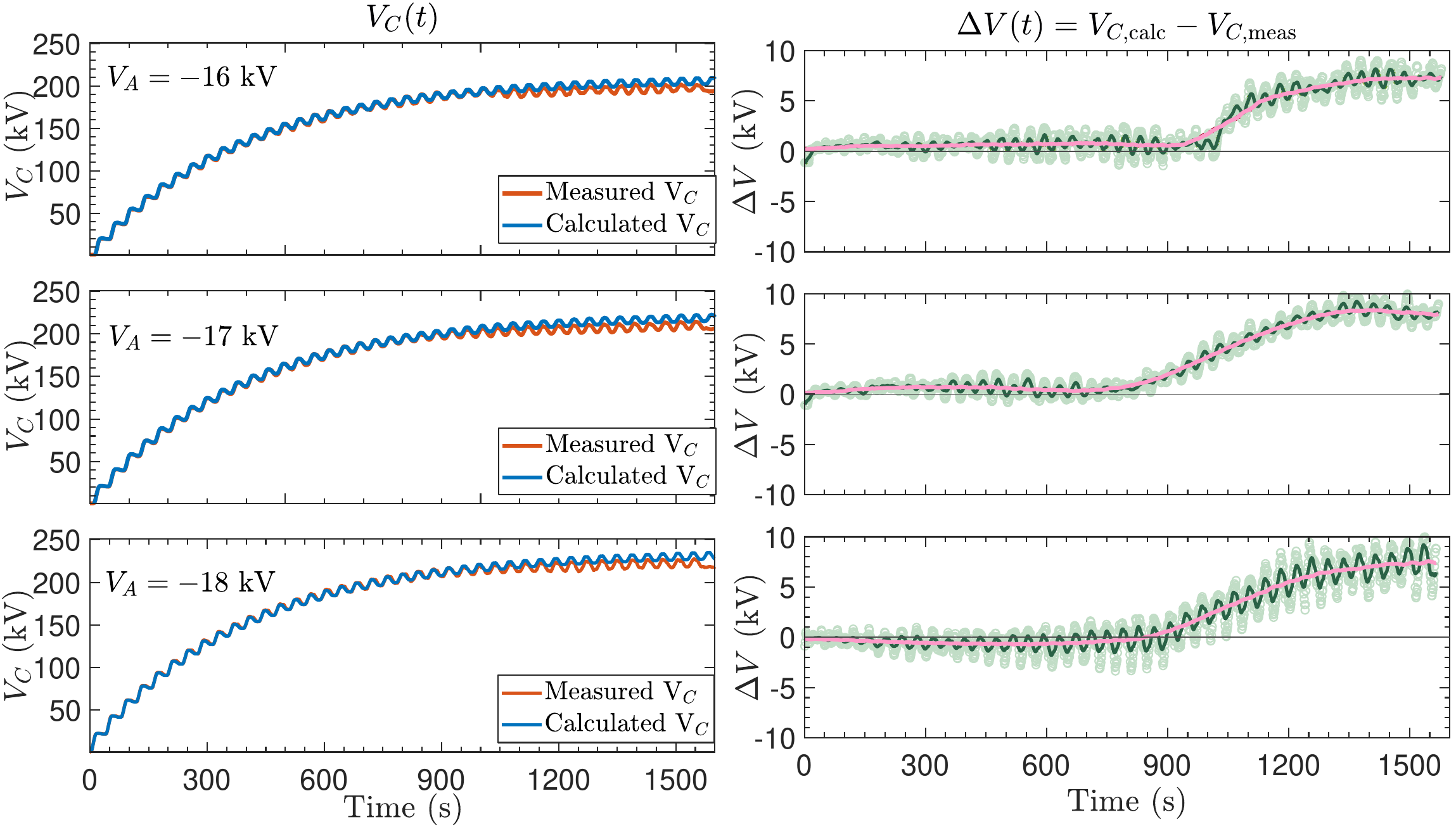}
    \caption{Comparison of measured and calculated (alignment-corrected) charging curves for higher input voltages, with rows corresponding to $V_A=-16$, $-17$, and $-18~\mathrm{kV}$. Left column: $C$ electrode voltage $V_C(t)$ from the capacitance model and from the measurement. Right column: difference $\Delta V(t)=V_{C,\mathrm{calc}}-V_{C,\mathrm{meas}}$, shown together with moving-average curves. For all three input voltages, the model reproduces the measured charging behavior well up to $V_C \approx 194-205~\mathrm{kV}$, above which a systematic positive difference develops, indicating the onset of an additional voltage-dependent charge-loss mechanism at high field.} \label{fig:charging_curves_delta_V_16-18kVneg}
\end{figure*}

Using these alignment-motivated capacitance values, with no further tuning, we then compare the model with the higher-voltage data for $V_A=-16$, $-17$, and $-18~\mathrm{kV}$ shown in Fig.~\ref{fig:charging_curves_delta_V_16-18kVneg}. For all three input voltages, the measured and calculated charging curves agree well up to $V_C \approx 200~\mathrm{kV}$. Above this value, however, the measured curves fall systematically below the model prediction, and the discrepancy grows rapidly as the charging curves approach saturation. The fact that this divergence begins at nearly the same value of $V_C$ for all three values of $V_A$ suggests that it is not primarily a consequence of an error in the assumed gain or input-voltage scaling of the model. Rather, it points to the onset of a voltage-dependent loss mechanism that becomes important once the $C$ electrode surface reaches a sufficiently high electric field.

\begin{figure*}[htb]
\centering
\includegraphics[width=1.00\linewidth]{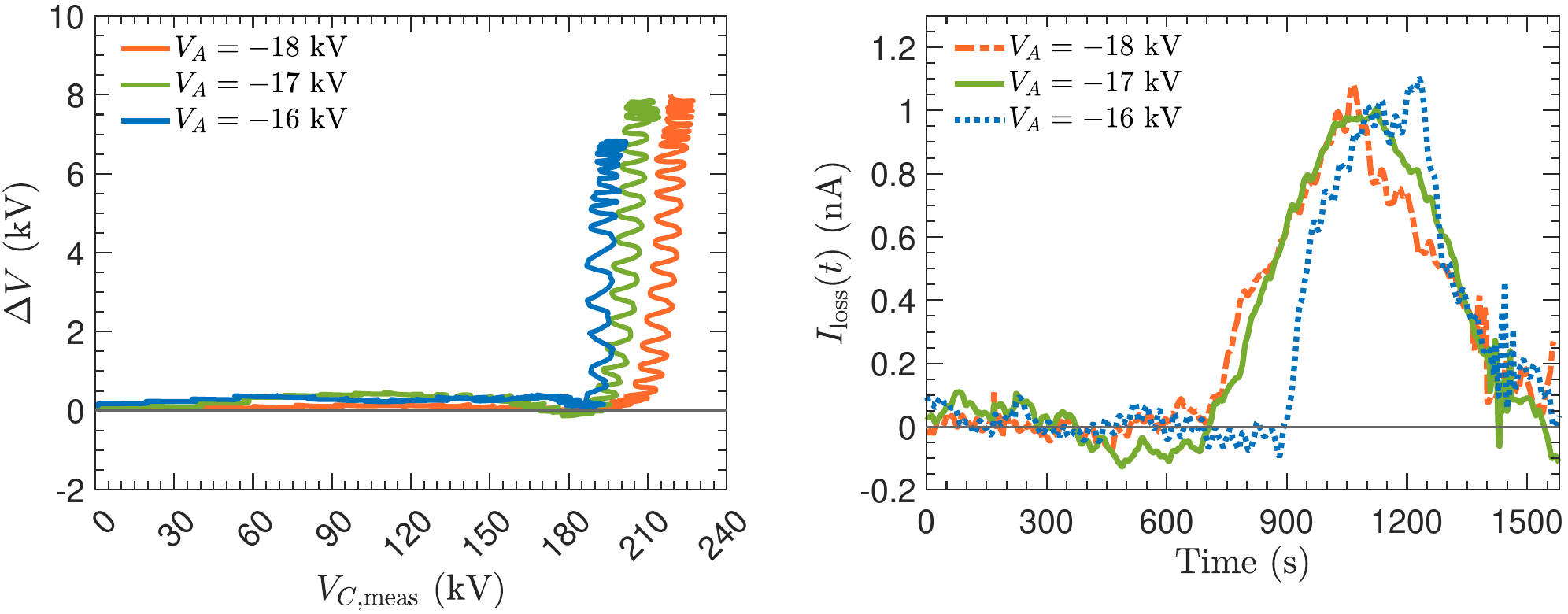}
    \caption{Residual-based analysis of the higher-voltage charging runs. Left: $\Delta V = V_{C,\mathrm{calc}} - V_{C,\mathrm{meas}}$ versus $V_{C,\mathrm{meas}}$ for $V_A=-16$, $-17$, and $-18~\mathrm{kV}$, smoothed with a moving-average of width 200. In all three cases, the voltage difference remains small until $V_{C,\mathrm{meas}} \approx 194$--$205~\mathrm{kV}$ and then rises rapidly, indicating the onset of an additional high-field loss channel. Right: corresponding effective loss current, $I_{\mathrm{loss}} = C_{CG}\, d(\Delta V)/dt$, plotted as a function of time. The onset times differ because the charging rates are not the same, but the peak currents are similar, reaching approximately $1.0$--$1.1~\mathrm{nA}$ in all three runs. Both the onset time and peak current are subject to the smoothing effect of the moving average filter.} \label{fig:current_18kVneg}
\end{figure*}

This interpretation is further supported by Fig.~\ref{fig:current_18kVneg}, where $\Delta V = V_{C,\mathrm{calc}} - V_{C,\mathrm{meas}}$ for $V_A=-16$, $-17$, and $-18~\mathrm{kV}$ is plotted as a function of $V_{C,\mathrm{meas}}$. The voltage difference remains close to zero throughout most of the charging cycle and then rises sharply once the $C$ electrode voltage exceeds about $194-205~\mathrm{kV}$, indicating a sudden onset of additional charge loss. If this discrepancy is attributed to dissipation from electrode~$C$, an effective loss current can be estimated from
\begin{equation}
\label{eq:I_loss}
\begin{aligned}
I_{\mathrm{loss}} = C_{\mathrm{CG}}\,\frac{d(\Delta V)}{dt}.
\end{aligned}
\end{equation}

The resulting currents, shown in the right-hand panel of Fig.~\ref{fig:current_18kVneg}, reach peak values of order $1.0~\mathrm{nA}$ for all three input voltages and then decrease. The subsequent decline is consistent with a transient high-field dissipation process that partially self-extinguishes, or with temporary conditioning of the emission or discharge site. The modest increase in the onset voltage between $V_A=-16~\mathrm{kV}$ and $-18~\mathrm{kV}$ is itself suggestive of a mild conditioning effect. In particular, the first clear onset of this behavior occurs in the $V_A=-16~\mathrm{kV}$ data, for which both $\Delta V$ and the inferred current rise more rapidly than in the $V_A=-17~\mathrm{kV}$ and $-18~\mathrm{kV}$ runs (see Figs.~\ref{fig:charging_curves_delta_V_16-18kVneg} and \ref{fig:current_18kVneg}). This may indicate that the initial high-field event partially modifies the active site and shifts the onset of the subsequent dissipation to slightly higher voltage. The magnitudes of the inferred currents are also very similar to those obtained from the short-timescale voltage-hold data shown in the left panel of Fig.~\ref{fig:voltage_hold_current_short_long_time}, as discussed in Sec.~\ref{sec:discuss-dissipation}, suggesting that the same field-activated dissipation process is responsible in both cases.

\begin{figure}[htb]
\centering
\includegraphics[width=1.00\linewidth]{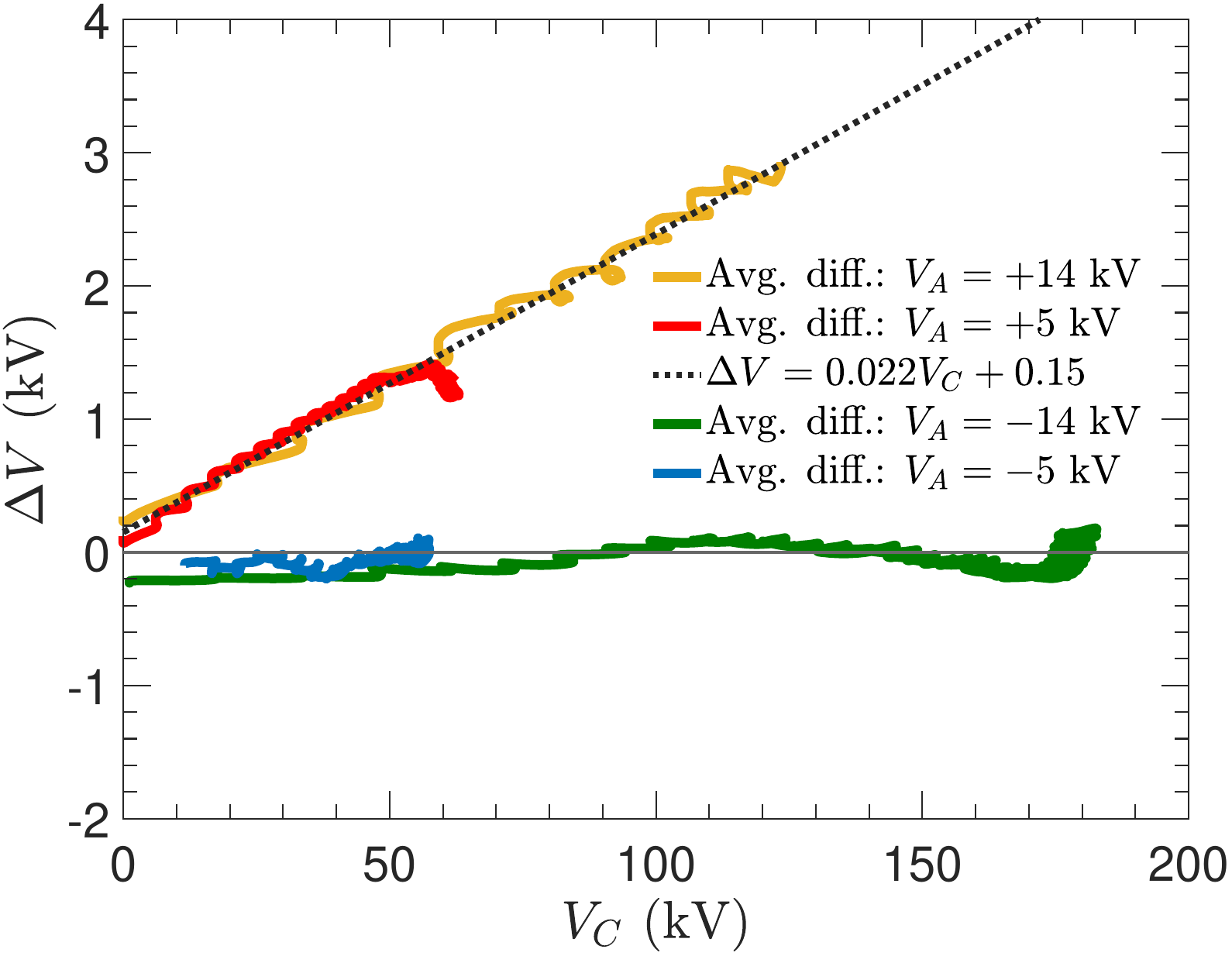}
    \caption{The voltage difference $\Delta V = V_{C,\mathrm{calc}} - V_{C,\mathrm{meas}}$ after applying the alignment-motivated capacitance corrections, shown for both bias polarities. For negative $V_A$ (corresponding to the $C$ electrode-voltage polarity used in the direct field-mill calibration), the difference remains close to zero. For positive $V_A$, the difference grows approximately linearly with $V_C$; the dashed line shows a linear fit used to estimate the additional sign-dependent calibration systematic for that polarity.} \label{fig:polarity_delta_V}
\end{figure}

Figure~\ref{fig:polarity_delta_V} compares the voltage differences after application of the alignment-motivated capacitance corrections for both bias polarities. For negative values of $V_A$, corresponding to the $C$ electrode voltage used in the direct field-mill calibration, the differences remain close to zero over the measured range. For positive values of $V_A$, however, the difference grows approximately linearly with $V_C$, with a near-zero intercept. This behavior suggests a polarity-dependent field-mill calibration slope rather than a residual deficiency in the capacitance model, as the same geometric corrections account for the opposite-polarity data. Speculatively, the triboelectric effect mentioned previously could be a contributor to this apparent polarity dependence. Because all direct calibration curves were acquired for one $C$ electrode-voltage polarity, the absolute voltage scale is only established for that polarity. The opposite-polarity charging curves demonstrate qualitative symmetry of the Cavallo charging process, but with an additional sign-dependent systematic uncertainty in the inferred $C$ electrode voltage.

A linear fit to the positive-polarity difference trend gives
\[
\Delta V_{\mathrm{pol}}(V_C) \approx 0.022\,V_C + 0.15,
\]
where $\Delta V_{\mathrm{pol}}$ is in kV and the coefficients are intended as an empirical description of the observed trend rather than as precise calibration constants. 
The fitted intercept is small compared with the voltage-dependent term, indicating that the polarity dependence appears primarily as a change in slope rather than as a fixed offset. The discrepancy extrapolates close to zero near $V_C=0$ and then grows approximately in proportion to $V_C$ over the measured range. For this reason, we use
\[
\delta V_{\mathrm{pol}}(V_C)=\left|\Delta V_{\mathrm{pol}}(V_C)\right|
\]
as an estimate of the additional sign-dependent systematic uncertainty for that polarity. At the maximum inferred $C$ electrode voltage of 255~kV, this corresponds to an added systematic of about $5.8~\mathrm{kV}$.

Overall, the comparison to the model supports a consistent picture of the system behavior. For the $C$-electrode polarity directly established by the field-mill calibration, the charging curves are quantitatively described by the capacitance model once realistic electrode alignment is taken into account. The opposite polarity exhibits similar charging dynamics, but carries an additional sign-dependent systematic uncertainty in the inferred absolute voltage scale. At higher voltages, the remaining discrepancy is not well explained by further geometric adjustments, but instead reflects the onset of a voltage-dependent charge-loss channel that limits the achievable $C$ electrode voltage in SF$_6$. Therefore, the measured saturation behavior can be viewed as the result of competition between the ideal Cavallo charging process and high-field dissipation, rather than as a breakdown of the electrostatic model itself.

\subsection{High-field charge dissipation and practical voltage limit in \texorpdfstring{SF$_6$}{SF6}}\label{sec:discuss-dissipation}

\begin{figure*}[htb]
\centering
\includegraphics[width=1.00\linewidth]{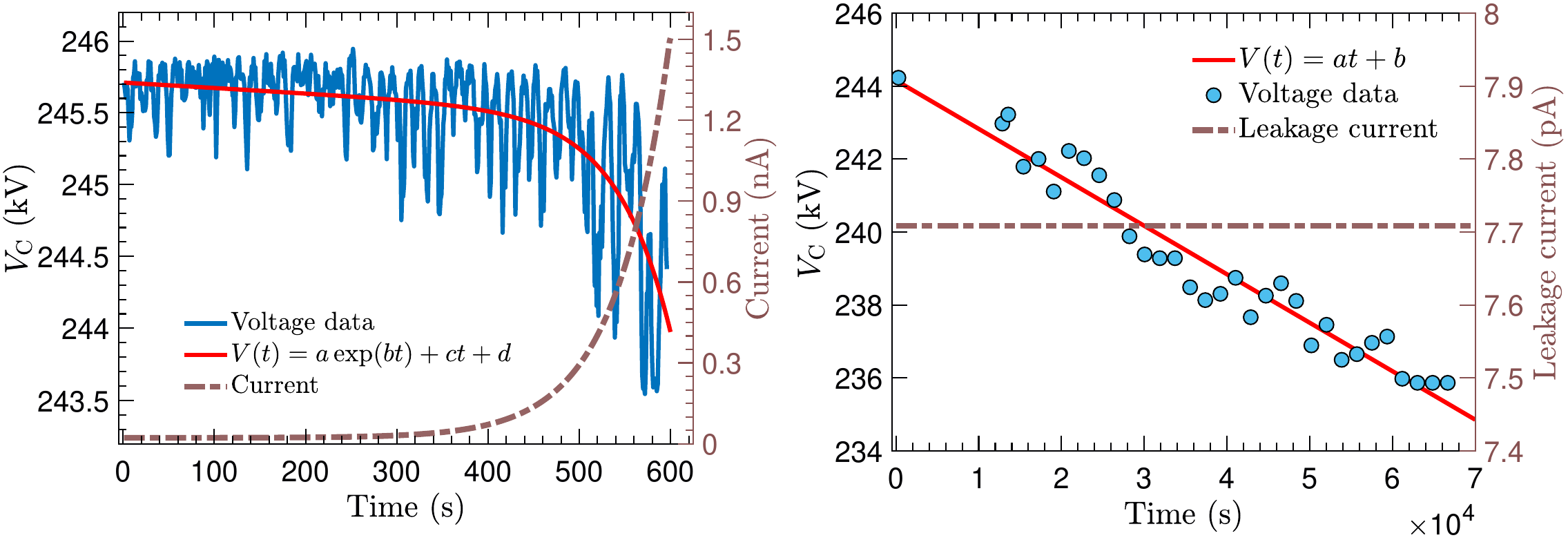}
    \caption{Left: short-timescale voltage-hold behavior of electrode~$C$ following charging to approximately 245~kV. The blue curve shows the measured voltage $V_{C}(t)$, and the red curve is a phenomenological fit to the decay. The brown curve, referenced to the right-hand axis, is the dissipation current inferred from the fitted voltage drop; the current is initially small ($\sim$~22 pA for $t<200$~s) and increases as the voltage approaches the onset of rapid discharge. Right: Long-timescale voltage-hold behavior of electrode~$C$. The measured voltage decreases approximately linearly over the holding interval, and the red line shows a linear fit to the data. The brown curve, referenced to the right-hand axis, gives the leakage current implied by the observed voltage decay and remains approximately constant at 7.7~pA, indicating stable pA-scale long-timescale dissipation.} \label{fig:voltage_hold_current_short_long_time}
\end{figure*}

The preceding discussion has focused on the accumulation of charge on the isolated $C$ electrode by means of the Cavallo multiplication process. Equally important, however, are the mechanisms by which this charge is lost once a large potential has been established. In the present apparatus, accumulation of a large charge on electrode~$C$ gives rise to high potentials and electric fields that can drive several dissipation pathways in parallel, including conduction through the bulk of the insulating support, surface leakage along the insulator, charge redistribution on nearby dielectric surfaces, and field-activated discharge processes in the surrounding medium. Charge loss can occur in two distinct modes: quasi-steady-state decay, a slow continuous process that gradually reduces the electrode potential over seconds, minutes, or hours; and transient discharge events, rapid processes that dissipate charge on microsecond or nanosecond time scales.  These processes are central to the determination of the long-term voltage stability of the system and, ultimately, the practical saturation voltage that can be reached. 

In our configuration, the $C$ electrode is supported by a PMMA insulating ring, and the expected bulk leakage through this support can be estimated from its geometry and resistivity. Reported values of the bulk resistivity of PMMA span several orders of magnitude, roughly $10^{14}$--$10^{19}~\Omega~\mathrm{cm}$, depending on material grade, surface condition, absorbed moisture, and measurement conditions. For the present order-of-magnitude estimate, we adopt a representative mid-range value of $\rho = 1\times10^{17}~\Omega~\mathrm{cm}$. Using $R=\rho L/A$ (where $L$ is the length and $A$ is the annular cross-section), the volume resistance of the standoff is approximately $2.5\times10^{16}~\Omega$. For a clean surface, the surface resistance may be of comparable order, approximately $1.9\times10^{16}~\Omega/\square$. At a voltage of order $255~\mathrm{kV}$, these values imply leakage currents at the level of at most tens of picoamperes (pA) and decay time constants exceeding $10^{5}~\mathrm{s}$. These estimates are only illustrative because, in practice, the dissipation may differ substantially from this idealized bulk-limited case, since the surface resistivity of PMMA is known to decrease by 5 to 6 orders of magnitude in the presence of adsorbed moisture or other contamination.~\cite{Baker1971}

The voltage-holding data show that the charge loss in the present apparatus is not well described by a single, voltage-independent leakage resistance. At $V_C=209~\mathrm{kV}$, no discernible voltage reduction was observed over a 15~min interval, indicating that the net dissipation was very small on that timescale ($< 6~\rm{pA}$ at 90\% confidence level based on a polynomial fit). At higher voltage, however, the behavior changes qualitatively. When the $C$ electrode was charged to approximately $245~\mathrm{kV}$, the voltage initially decreased from about $245.7~\mathrm{kV}$ to approximately $244~\mathrm{kV}$ within roughly 10~min, as shown in Fig.~\ref{fig:voltage_hold_current_short_long_time} (left panel). At later times, the decay slowed substantially. Over a holding time of approximately 19~h, the voltage decreased further to about $236~\mathrm{kV}$, corresponding to a total drop of $3.7\%$ and to an average long-timescale leakage current of $7.7~\mathrm{pA}$, as shown in Fig.~\ref{fig:voltage_hold_current_short_long_time} (right panel). The effective resistance implied by this long-timescale decay is of order $3\times10^{16}~\Omega$, which is of the same order as the representative resistance estimated for the PMMA support. This agreement should be regarded as an order-of-magnitude consistency check rather than evidence of the dominant leakage path, since pA-level currents may include comparable contributions from PMMA bulk conduction and insulator surface leakage.  Nevertheless, the long-timescale holding data show that the apparatus can maintain a charged electrode in a stable quasi-steady leakage regime at the pA level, but they do not determine which weak leakage pathway dominates under these conditions.  As such, it is important to note that the fits shown in Fig.~\ref{fig:voltage_hold_current_short_long_time} are phenomenological descriptions of the total observed voltage loss, incorporating the combined effect of all leakage and dissipation pathways, rather than unique physical models of any particular loss mechanism.

This quasi-steady pA-scale regime is clearly distinct from the short-timescale behavior near $245~\mathrm{kV}$, which requires an additional field-activated loss channel.  As shown in the left panel of Fig.~\ref{fig:voltage_hold_current_short_long_time}, the inferred dissipation current is initially small, at the level of tens of picoamperes ($\sim 22~\mathrm{pA}$ for $t<200~\mathrm{s}$), but then rises rapidly with the onset of the transient discharge process. Near the end of the short-timescale hold interval, the effective current exceeds $1~\mathrm{nA}$, which is more than two orders of magnitude larger than the current inferred from the long-timescale decay. The onset of the transient loss occurs while $V_C$ is already decreasing slightly, indicating that the instability is not determined solely by the instantaneous macroscopic $C$ voltage. Instead, the data suggest a finite incubation time or activation time of the loss channel.  This is consistent with the growth of a local discharge precursor---such as activation of a microscopic emission site or dielectric charge redistribution---that continues to develop briefly even as the $C$ voltage relaxes.

Although the short-timescale trace does not extend far enough to show the subsequent relaxation directly, the long-timescale hold data in the right panel of Fig.~\ref{fig:voltage_hold_current_short_long_time} imply that this elevated current must decay rapidly: had it remained at the nA level, the $C$ electrode voltage at the second long-timescale point would have been far below the observed value. This shows that the short-timescale voltage loss arises from a transient, self-extinguishing dissipation process rather than from the approximately ohmic leakage responsible for the later slow decay. In this picture, the long-timescale holding data characterize the intrinsic stability of the charged electrode, whereas the rapid late-time decay reflects a transient, field-activated dissipation process superimposed on the much smaller quasi-steady leakage current.

This interpretation is also consistent with the charging curves presented in Figs.~\ref{fig:charging-curves-1} and \ref{fig:charging-curves-2}. Although the capacitance model and the nominal electrode geometry imply that higher values of $V_C$ should in principle be obtainable, the measured charging curves saturate near 250~kV because the charge delivered to electrode $C$ during each Cavallo cycle is increasingly offset by voltage-dependent charge loss. This implies that the observed saturation voltage in SF$_6$ reflects an equilibrium between charge transfer and charge dissipation, rather than the ideal geometry-limited gain of the multiplier. This distinction is important, because it shows that the maximum voltage achieved in room-temperature commissioning depends not only on the electrostatic design of the apparatus, but also on the microscopic processes responsible for loss and discharge at high field. 

The present measurements do not identify the microscopic origin of the transient loss channel. Possible contributors include surface conduction along the PMMA ring, localized ionization in the SF$_6$, and microdischarge initiated at regions of enhanced electric field on rough electrode surfaces or at metal-insulator-gas (triple junction) interfaces.\cite{Schueller2015} Regardless of the precise mechanism, the strong voltage dependence of the dissipation and its close correlation with the maximum achievable voltage indicate that surface and interface phenomena are the dominant practical limitation in the present measurements. This conclusion is also consistent with the observed sensitivity of the apparatus to electrode surface condition, including the performance degradation following discharge and the recovery obtained after repolishing.


\subsection{Voltage performance in SF$_6$ and implications for cryogenic operation}\label{sec:discuss-sf6-limitations}

The maximum field-mill-inferred $C$-electrode voltage observed in SF$_6$ was approximately $255~\mathrm{kV}$. For this polarity, the inferred absolute voltage scale carries the additional sign-dependent systematic discussed in Secs.~\ref{sec:calibration} and \ref{sec:discuss-comparision}, estimated from the linear trend in Fig.~\ref{fig:polarity_delta_V} to be about $5.8~\mathrm{kV}$ at the maximum inferred voltage. The corresponding quasi-uniform field between electrodes~$C$ and~$D$ is approximately $33~\mathrm{kV/cm}$, and the maximum field on the surface of electrode~$C$ inferred from the COMSOL field map is approximately $45~\mathrm{kV/cm}$;\cite{Blatnik2026} both values carry the same fractional scale uncertainty. These values are still substantially lower than the experimentally established Paschen breakdown strengths for SF$_6$, which are typically between $70$ and $85~\mathrm{kV/cm}$ for the specific pressure and gap distance of the present experiment.\cite{Malik1979, Kind1974} However, the applicability of Paschen's law is limited to approximately uniform electric fields and to low pressure--distance products, $pd < 2~$bar cm. Outside this regime, breakdown may proceed via the streamer mechanism rather than the Townsend mechanism. For the streamer regime at elevated pressures, Kuffel~\textit{et al.}\cite{kuffel2000} give the empirical relation
\[
V_b(\mathrm{kV}) = 40 + 68(pd),
\]
where $p$ is in bar and $d$ is in cm. For the experimental conditions $p = 0.80~\mathrm{bar}$ and $d = 7.62~\mathrm{cm}$, this relation predicts a breakdown voltage of $455~\mathrm{kV}$, which is still substantially higher than the approximately $255~\mathrm{kV}$ observed in the present study.

The reason for this disparity is that in practical high-voltage systems, the limiting factor for insulation performance is often not the intrinsic dielectric strength of the medium, but rather phenomena occurring at the electrode-gas interface.\cite{Malik1979, kuffel2000, Pedersen1975} In experimental setups with inherently non-uniform electric fields, the breakdown characteristics of SF$_6$ are more complex than in uniform fields. In the present experiment, field non-uniformity arises from two distinct sources: the macroscopic geometry of the electrodes and the microscopic surface roughness of the electrodes, particularly electrode $C$. Microscopic asperities on the electrode surface can produce strong local field enhancement and play an important role in the breakdown process. Electropolishing is relevant in this context because it preferentially rounds the tips of such asperities, where the local electric field is highest, and thus reduces the microscopic field enhancement relative to the local macroscopic field.  This is commonly expressed through an enhancement factor $\beta$ defined by $E_{\mathrm{micro}}=\beta E_{\mathrm{macro}}$. Therefore, a non-electropolished surface can retain sharper protrusions and larger local values of $\beta$, making field emission, local ionization, and discharge initiation more likely at a lower macroscopic field.

Previous work indicates that in SF$_6$, when the product of the gas pressure, $p$, and the height of a surface protrusion, $R$, exceeds a critical threshold of approximately 40 bar$~\mu$m,~\cite{Pedersen1975} the dielectric breakdown strength can deviate significantly from the value implied by the ideal Paschen curve. This is attributed to the high gas density, which enables the rapid generation of ionization charge sufficient to initiate a streamer within the intense local electric field near microscopic electrode-surface asperities. Consequently, the macroscopic breakdown of the entire inter-electrode gap is triggered by a localized, microscopic event. Because electrode $C$ was not electropolished, its inherent surface roughness likely contributed to the lower-than-expected breakdown voltage.

Consequently, the maximum voltage reached in approximately 600 Torr of SF$_6$ may not represent the intrinsic limit of the Cavallo geometry. Rather, the present data indicate that room-temperature gas operation is constrained by a combination of nonuniform fields, electrode surface condition, and discharge initiation at localized microscopic asperities, particularly on electrode $C$. In this sense, the SF$_6$ measurements provide a conservative demonstration of the voltage multiplication process and of the associated charging dynamics, while the ultimate performance of the apparatus in a cryogenic dielectric medium is expected to depend primarily on the control of interfacial breakdown phenomena rather than on the ideal bulk dielectric strength alone.

Liquid nitrogen (LN$_2$) provides a useful next-stage test medium because it is cryogenic and considerably closer to the intended operating environment than room-temperature gas, while remaining experimentally simpler to operate than liquid helium (LHe). Since the present SF$_6$ data reached a $C$-electrode voltage of approximately 255~kV, corresponding to an approximately $33~\mathrm{kV/cm}$ quasi-uniform field and an approximately $45~\mathrm{kV/cm}$ maximum field on electrode~$C$, the observed limitation does not by itself suggest a fundamental ceiling for operation in LN$_2$. These field estimates inherit the same sign-dependent scale uncertainty discussed above, but they nevertheless remain well below values that would imply an intrinsic geometry-limited ceiling at this stage. One expects the denser liquid dielectric to provide greater breakdown margin than the gas phase, provided that the electrode surfaces remain smooth and that bubble formation is suppressed.\cite{Sauers2011, Maeda1998} At the same time, the relevant limitation in LN$_2$ may not be the nominal bulk dielectric strength of the liquid, but rather local heating, field emission, trapped gas, impurities, or bubble nucleation at rough surfaces and metal-insulator-liquid interfaces.\cite{Maeda1998}

As a result, LN$_2$ operation should be viewed not merely as a simple extrapolation from the SF$_6$ data, but as a test of whether the practical voltage ceiling shifts upward toward the value implied by the capacitance model and the designed cryogenic gain.  The maximum sustainable $V_C$, the reproducibility of the charging curves near saturation, the long-term hold stability, and the extent to which transient discharge events are suppressed relative to operation in SF$_6$ are important gauges. Improvement in these quantities would support the interpretation that the present room-temperature limitation is dominated by gas-phase and surface-triggered discharge processes rather than by the multiplier geometry itself.

The ultimate intended operating medium, however, is superfluid $^4$He. Previous studies indicate that electrical breakdown in liquid helium depends sensitively on pressure, electrode surface condition, and stressed area, and is consistent with a surface-initiated process closely correlated with Fowler--Nordheim field emission from cathode asperities.\cite{Phan2021} Measurements in a medium-scale apparatus at $0.4~\mathrm{K}$ further demonstrated that fields exceeding $100~\mathrm{kV/cm}$ can be sustained in a $1~\mathrm{cm}$ gap between electropolished stainless-steel electrodes over a wide pressure range.\cite{Ito2016} For the cryogenic nEDM application, the required operating point is $75~\mathrm{kV/cm}$, corresponding to $635~\mathrm{kV}$, and the adoption of a pressurized He-II environment is intended to reduce breakdown probability by suppressing bubble formation.\cite{Ito2026}  In this context, the Cavallo multiplier is advantageous because it removes the direct thermal connection between the high-voltage electrode and an external feed system. Taken together, these considerations suggest that the present room-temperature demonstration does not indicate an intrinsic geometry-limited barrier to eventual operation in LHe.


\section{Summary and conclusion}\label{sec:conclusion}

We have demonstrated room-temperature operation of a Cavallo multiplier built for eventual cryogenic use. In approximately $600~\mathrm{Torr}$ of SF$_6$, the system reached $V_C=(237.7\pm1.7)~\mathrm{kV}$ for the directly calibrated polarity.  The measured charging curves exhibit the expected behavior of progressive charge accumulation on the isolated $C$ electrode, and their overall form is consistent with the capacitance-based description of the multiplier once practical effects such as electrode alignment are taken into account. For the other polarity, output voltages up to approximately 255~kV were reached in SF$_6$ with a 25~kV bias voltage.  This polarity carries an additional sign-dependent calibration systematic, estimated to be about $5.8~\mathrm{kV}$ at the maximum voltage.

The measurements further show that the principal limitations of the present room-temperature operation arise not from the multiplication mechanism itself, but from discharge processes and charge-loss pathways associated with the electrode surfaces and insulating interfaces. In particular, the observed sensitivity to post-discharge surface condition, together with the restoration of performance following repolishing, indicates that localized field enhancement at surface imperfections is a dominant factor in setting the achievable voltage in SF$_6$. Voltage-hold measurements additionally show that, away from the onset of discharge, the long-timescale leakage current is at the picoamp level, demonstrating that the apparatus is capable of maintaining a highly charged electrode with only modest dissipation.

These results provide an important validation of the Cavallo multiplier concept as an \textit{in situ}, low-current high-voltage source for experiments in which conventional feedthrough-based delivery is difficult or undesirable. More broadly, the present study indicates that electrode surface condition, alignment, and control of insulation-related charge dissipation are the key practical requirements for reliable operation. Because the dielectric limitations encountered in room-temperature SF$_6$ are not expected to represent the intrinsic performance of the apparatus in a cryogenic liquid, an important next step for the further development of this approach is evaluation in LN$_2$, where the attainable voltage, hold stability, and breakdown behavior can be assessed under conditions more closely matched to the intended experimental application.


\begin{acknowledgments}
This work was supported by the United States Department of Energy,
Office of Science, Office of Nuclear Physics through Los Alamos
National Laboratory under Contract No. 89233218CNA000001 via proposal 2023LANLEED3, the Los Alamos National Laboratory Integrated Contract Order No.
4000129433 with Oak Ridge National Laboratory, and the National Science Foundation Grants
NSF-2110898, NSF-1822515, and NSF-1812340—“Fundamental Studies in Nuclear Physics.”

\end{acknowledgments}




\section*{DATA AVAILABILITY}
The data that support the findings of this study are available from the corresponding author upon reasonable request.  


\section*{Generative AI Statement}
The authors acknowledge the use of generative artificial intelligence tools, specifically GPT~5.4 and 5.5 and Gemini~3.1 Flash, to assist with literature research and language editing, including improvement of prose flow, grammatical correctness, and formal academic tone. All AI-assisted output was reviewed, verified, and, where necessary, revised by the authors for accuracy, completeness, and consistency with the scientific content of the manuscript. These tools were not used as substitutes for scientific judgment. The authors accept full responsibility for the integrity and accuracy of the final manuscript.


\section*{References}
\bibliography{biblio}

\end{document}